\documentclass[iop]{emulateapj}
\usepackage{amsmath}
\usepackage{natbib}

\submitted{accepted by the Astrophysical Journal}

\usepackage{soul}
\usepackage{color}
\usepackage{textcomp}

\usepackage{graphicx}
\usepackage{hyperref}

\newcommand{\rxte}{\emph{RXTE}}

\newcommand{\chandra}{\emph{Chandra}}

\newcommand{\xmm}{\emph{XMM-Newton}}

\newcommand{\wsim}{\ensuremath{\sim}}

\newcommand{\msun}{$\rm M_{\odot}$}

\newcommand {\HS}{H1743$-$322}
\newcommand{\GX}{GX~339$-$4}
\newcommand{\FF}{XTE~J1550$-$564}
\newcommand{\SF}{GRO~J1655$-$40}
\newcommand{\FU}{4U~1543$-$47}
\newcommand{\ST}{XTE~J1752$-$223}
\newcommand{\STT}{XTE~J1720$-$318}

\begin{document}


\title{Complete multiwavelength evolution of Galactic black hole transients during outburst decay I: conditions for ``compact" jet formation}

\author{E. Kalemci\altaffilmark{1},
        T. Din\c{c}er\altaffilmark{1},
        J. A. Tomsick\altaffilmark{2},
        M. M. Buxton\altaffilmark{3},
        C. D. Bailyn\altaffilmark{3},
        Y. Y. Chun\altaffilmark{1}
}

\altaffiltext{1}{Faculty of Engineering and Natural Sciences, Sabanc\i\ University, Orhanl\i-Tuzla, 34956, Istanbul, Turkey}

\altaffiltext{2}{Space Sciences Laboratory, 7 Gauss Way, University of
California, Berkeley, CA, 94720-7450, USA}

\altaffiltext{3}{Astronomy Department, Yale University, P.O. Box 208101, New Haven, CT 06520-8101, USA}



\begin{abstract}

Compact, steady jets are observed in the near infrared and radio bands in the hard state of Galactic black hole transients as their luminosity decreases and the source moves towards a quiescent state. Recent radio observations indicate that the jets turn off completely in the soft state, therefore multiwavelength monitoring of black hole transients are essential to probe the formation of jets. In this work we conducted a systematic study of all black hole transients with near infrared and radio coverage during their outburst decays. We characterized the timescales of changes in X-ray spectral and temporal properties and also in near infrared and/or in radio emission. We confirmed that state transitions occur in black hole transients at a very similar fraction of their respective Eddington luminosities. We also found that the near infrared flux increase that could be due to the formation of a compact jet is delayed by a time period of days with respect to the formation of a corona.  Finally, we found a threshold disk Eddington luminosity fraction for the compact jets to form. We explain these results with a model such that the increase in the near infrared flux corresponds to a transition from a patchy, small scale height corona along with an optically thin outflow to a large scale height corona that allows for collimation of a steady compact jet. We discuss the timescale of jet formation in terms of transport of magnetic fields from the outer parts of the disk, and also consider two alternative explanations for the multiwavelength emission: hot inner accretion flows and irradiation.

\end{abstract}




\section{Introduction}\label{sec:intro}

Galactic black hole transients (GBHT) show distinct spectral and temporal changes during the decay of outbursts across all wavelengths.   At the start of the outburst decay, the GBHTs are usually in the soft state, the X-ray spectrum is dominated by emission from an optically thick, geometrically thin disk, and the rms amplitude of variability is less than a few percent. As the flux decays, a sudden increase occurs in the rms amplitude of variability accompanied by an increase in the non-thermal emission often associated with Compton scattering of soft photons by a hot electron corona. After $\sim$10--20 days or less, the non-thermal emission (often modeled with a power-law in the X-ray spectrum) dominates the X-ray flux above 3 keV as the source enters the hard state. The detailed description of spectral states and the general evolution of GBHTs during the entire outburst can be found in \cite{Belloni10_jp} and references therein. The changes in X-ray spectral and temporal properties specifically during the outburst decay are described in detail in \cite{Kalemci03}.

Contemporaneous observations of GBHTs in radio, optical and near infrared (NIR) during the decay provide additional information about the accretion/ejection behavior of these sources. Radio observations track the behavior of jets in these systems. It is well established that the jet is quenched significantly in the soft state \citep{Corbel02, Russell11}, and a steady compact jet is observed in the hard state during outburst decay as evidenced by a flat to inverted radio spectrum \citep{Corbel00, Fender01b}. Jets may also be revealing themselves when secondary maxima in the optical and NIR fluxes occur during decay. At the start of the decay, the NIR fluxes decay along with the X-ray flux. At some point, the NIR flux increases, peaks and then falls down within a timescale of months as shown in Fig.~\ref{fig:harden1}.  This happens when the source is fully back in the hard state with its X-ray spectrum close to its hardest level \citep{Kalemci05, Coriat09, Russell10, Dincer12, Buxton12}.The spectral energy distributions (SED) created from data during the NIR peaks of \FU\ \citep{Buxton04, Kalemci05} and \FF\ \citep{Jain01_b, Russell10} show a flat or inverted power-law at radio frequencies that breaks, usually at NIR wavelengths, to a second power-law with negative spectral index consistent with emission from a compact, conical jet \citep{Blandford79, Hjellming88}. Given the similar NIR evolution observed in \GX\ \citep{Coriat09,Buxton12} and partially also in \ST\ \citep{Chun13}, it is reasonable to assume that the NIR peaks in the hard state have a jet origin. On the other hand, SEDs created during the early parts of the NIR peak for \GX\ are not consistent with optically thin emission from a jet: they are rather flat, even inverted up to the $V$ band. This can be explained with extra emission components at high frequencies on top of the the optically thin synchrotron \citep{Coriat09, Dincer12, Rahoui12}.  

Compared to the NIR coverage, the radio coverage of GBHTs is usually sparse (see Figures~\ref{fig:harden1} and \ref{fig:harden2}).  During the transition from the soft state to the hard state, few radio detections exist  for \STT \citep{Brocksopp05,Fender09}, and \HS \citep{Jonker10, MillerJ12}. For two cases, there is enough simultaneous coverage of radio, NIR-optical, and X-rays which provided a better understanding of the relation between the jet and the NIR peak. Observations of GX~339$-$4 published recently \citep{Corbel12, Corbel13} show that the NIR rise may correspond to a transition in radio from optically thin to optically thick emission in the 2011 outburst decay. Similarly,  \ST\ also shows an increase in the radio spectral index  $\alpha$ (defined as $F_{\nu} \; \propto \; \nu^{\alpha}$ where $F_{\nu}$ is the radio flux density and $\nu$ is the frequency), and radio spectrum becoming consistent with a flat spectrum during a large NIR peak during the decay of the outburst  \citep[see Fig.~\ref{fig:harden2}, and also ][]{Brocksopp13,Chun13}. 

The NIR peaks that are associated with jets also exist in the hard state during the rise of the outbursts \citep{Coriat09, Buxton12, Russell07}. However, it is difficult to catch the start of the outbursts, and often, when the source is detected in X-rays, the compact jets are already present. On the other hand, the outburst decays allow us to investigate the relation between the X-ray spectral properties and the NIR/radio flux and spectral evolution in detail as the multiwavelength jet emission turns on and increases while the GBHTs make a transition from the soft to the hard state.The properties of some of the individual black hole sources are already discussed in \cite{Kalemci05, Kalemci06, Jonker10, Ratti12}, and preliminary analysis of the overall behavior of many sources  is discussed in \cite{Kalemci06_mqw, Kalemci08_kol}. In this work, we present an in-depth, systematic multiwavelength analysis of all GBHTs covered well with \rxte\ in X-rays, SMARTS \citep{Subsavage10}, in NIR and in radio. The failed outbursts are excluded from this study because we are interested in sources that go through the soft state. We emphasize changes (or lack thereof) in X-ray spectral properties when the jet related emission is first observed in NIR and radio, and we  also discuss the timescale for jet formation.  A second article which will discuss the relation between jet emission and X-ray timing properties is also being prepared \citep{Dincer13}.




\section{Observations and Analysis}\label{sec:obs}

\subsection{X-ray spectral analysis}\label{subsec:rxtespec}

 We use PCA in the 3--25 keV band and HEXTE in the 15--200 keV band (see \citealt{Bradt93} for instrument descriptions) and fit the spectra together.  However, we do not include the HEXTE data if the background-subtracted 20--100 keV count rate in cluster A is less than 3 cts/s.  Also, HEXTE data were not used after cluster B stopped rocking on December 14, 2009.  For PCA, we use all the available PCUs for all observations and include systematic errors at a level of 0.8\% up to 7 keV and 0.4\% above 7 keV.

For all sources, the HEXTE background fields are checked using the \emph{HEXTEROCK} utility and compared to Galactic bulge scans. Only fields not contaminated with sources or strong background are used. HEXTE spectra are corrected for deadtime \citep{Rothschild98}. 

The Galactic ridge emission is a factor for some of the sources at low flux levels. The ridge contribution is determined by one of the two methods described below. If there is a simultaneous observation at low flux levels with an X-ray telescope (such as \xmm\, or the \chandra) along with \rxte, we compare the spectra to determine the ridge spectrum. If there is no such simultaneous observation, we check the PCA light curves at the lowest flux levels and look for a level of constant flux. These observations are merged to model the ridge spectrum. 

The first spectral model we try for all observations consists of absorption (``phabs'' in XSPEC), a smeared edge \citep[``smedge'' in XSPEC,][]{Ebisawa94}, a multicolor disk blackbody \citep[``diskbb'' in XSPEC,][]{Makishima86}, a power law ``power'' in XSPEC), a narrow Gaussian to model the iron line, and, if necessary, the ridge emission. This model has been commonly used for the spectral analysis of black holes in the hard state \citep{Tomsick00,Sobczak00,Kalemci05}.  The hydrogen column density is fixed to values found in the literature. The smeared edge width is fixed to 10 keV. For each observation, we then introduce a high energy cut-off (``highecut'' in XSPEC) to the model and check the improvement in the fit using F-test. If the chance probability becomes less than 0.001, we include the cut-off in the fit. 

\begin{table}[h]
\caption{\label{table:md} Masses and distances used}
\begin{tabular}{l|c|c|c|c} \hline \hline
Source & Mass & Distance & b\_sep\footnote{Binary seperation, in lightseconds} &References\footnote{1: \cite{Dunn10}, 2: \cite{Foellmi06}, 3: \cite{Orosz11},  4: \cite{Steiner12}, 5: \cite{Shaposhnikov10}, 6: \cite{Chun13}, 7: \cite{CadolleBel04}.  Most of the references are from \cite{Dunn10}, Table 1.} \\
       & (\msun) & kpc &  & \\ \hline
\SF & 7.0$\pm$0.2 & 3.2$\pm$0.2\footnote{\cite{Foellmi06} reported an alternative distance of $<$1.7 kpc}& 38 & 1, 2 \\
\GX & [8]\footnote{Based on \cite{Kreidberg12} and \cite{Ozel10}} & 8$\pm$2 & 42.1 & 1 \\
\FF & 9.1$\pm$0.6 & 4.4$\pm$0.5 & 37 & 3 \\
\HS & [8] & 8.5$\pm$0.8 & - & 4 \\
\FU & 9.4$\pm$2 & 7.5$\pm$0.5 & 23 & 1 \\
\ST & 9.5$\pm$1.5 & [8]\footnote{3.5$\pm$1.5 according to \cite{Shaposhnikov10}, however our work indicates distance greater than 5 kpc \citep{Chun13}} & - & 5, 6 \\
XTE J1720$-$318 & [8] & [8] & - & 7 \\ \hline
\end{tabular}
\end{table}

Using the black hole mass and distance values reported in Table~\ref{table:md}, we calculated the Eddington Flux for each source. If necessary, we extrapolated our X-ray spectra to the 3--200 keV band using the spectral fits and defined the Eddington Luminosity Fraction (ELF) as the ratio of 3--200 keV flux to the Eddington Flux of each source. This method underestimates the actual Eddington Luminosity Fraction because we are not calculating the bolometric luminosity. There are also some uncertainties coming from the extrapolation of the X-ray spectrum to 3--200 keV when HEXTE is not used. However, given the large uncertainties in mass and distance and also since the total energy budget should be dominated by X-rays, these uncertainties do not affect our results significantly.

\subsection{X-ray temporal analysis}\label{subsec:rxtetim}

We use T\"{u}bingen Timing Tools in IDL to compute the power density spectra (PDS) of all observations using PCA light curves in the 3--25 keV band. The dead time effects are removed according to \cite{Zhang95} with a dead-time of $\rm 10\,\mu s$ per event, and the PDS is normalized according to \cite{Miyamoto89}. Broad and narrow Lorentzians are used for fitting \citep{Kalemci05,Pottschmidt02th}. The rms amplitudes are calculated by integrating Miyomoto normalized PDS from 0 Hz to infinity. The rms amplitudes are corrected with a factor ${T}\over{(T-(R+B))}$, where $T$ is the overall count rate, $B$ is the background rate, and $R$ is the count rate due to the Galactic ridge, to obtain the variability inherent to the source \citep{Berger94}\footnote{In \cite{Kalemci06}, the correction is erroneously stated with the square of the factors, which is the correct factor for the PSD but not for the rms amplitude of the variability.}. In this work, the timing information is only used for determining the state transitions.  For a detailed analysis of timing properties of all sources during the decay, see \cite{Dincer13}

\epsscale{1.25}
\begin{figure}[b]
\plotone{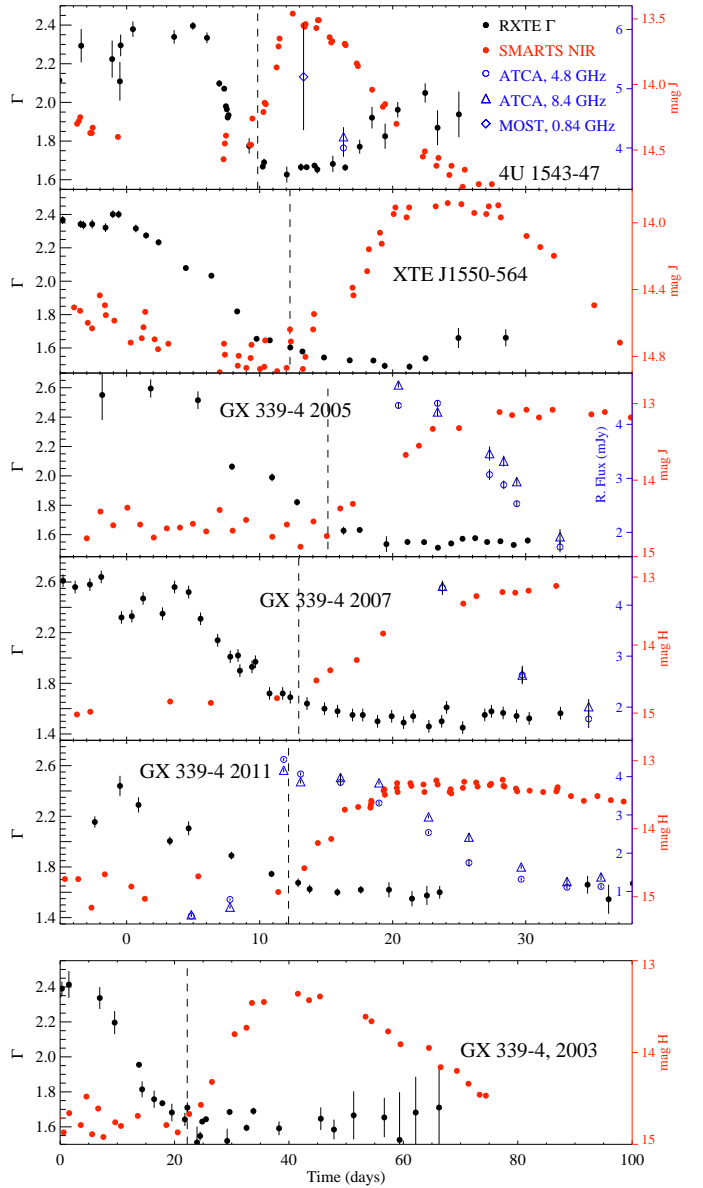}
\caption{\label{fig:harden1}
Evolution of power law photon indices ($\Gamma$, black solid circles) along with evolution in near infrared magnitudes (gray solid circles, red in the color version), and radio fluxes (open circles and triangles, blue in the color version). The radio fluxes (in units of mJy) are indicated in the inner part of the y-axis on the right hand side. The dashed lines show the start of the NIR flare. Time 0 denotes the time of the timing transition (see \S~\ref{sec:results} and the Appendix). A color version of this figure is present in the online version.
}
\end{figure}
\epsscale{1.0}

\subsection{Determining NIR and Radio transitions}\label{subsec:multi}

An important goal of this work is to find the time of compact jet formation relative to changes in X-ray spectral and temporal parameters. We assume that the flare in the NIR is due to the emission from a compact jet (see Fig.~\ref{fig:harden1}). For an in-depth discussion of this assumption, see \S{\ref{subsec:rebright}}. The NIR flare rises above a baseline NIR emission, which may arise from the accretion disk. We fit the baseline NIR flux as a function of time with an exponential to determine the non-flare emission. Then, we fit the initial rise of the flare above this baseline with a linear function to find the start time for \GX, \FU, and \FF. This procedure is explained in detail in the Appendix.

\begin{figure}[t]
\plotone{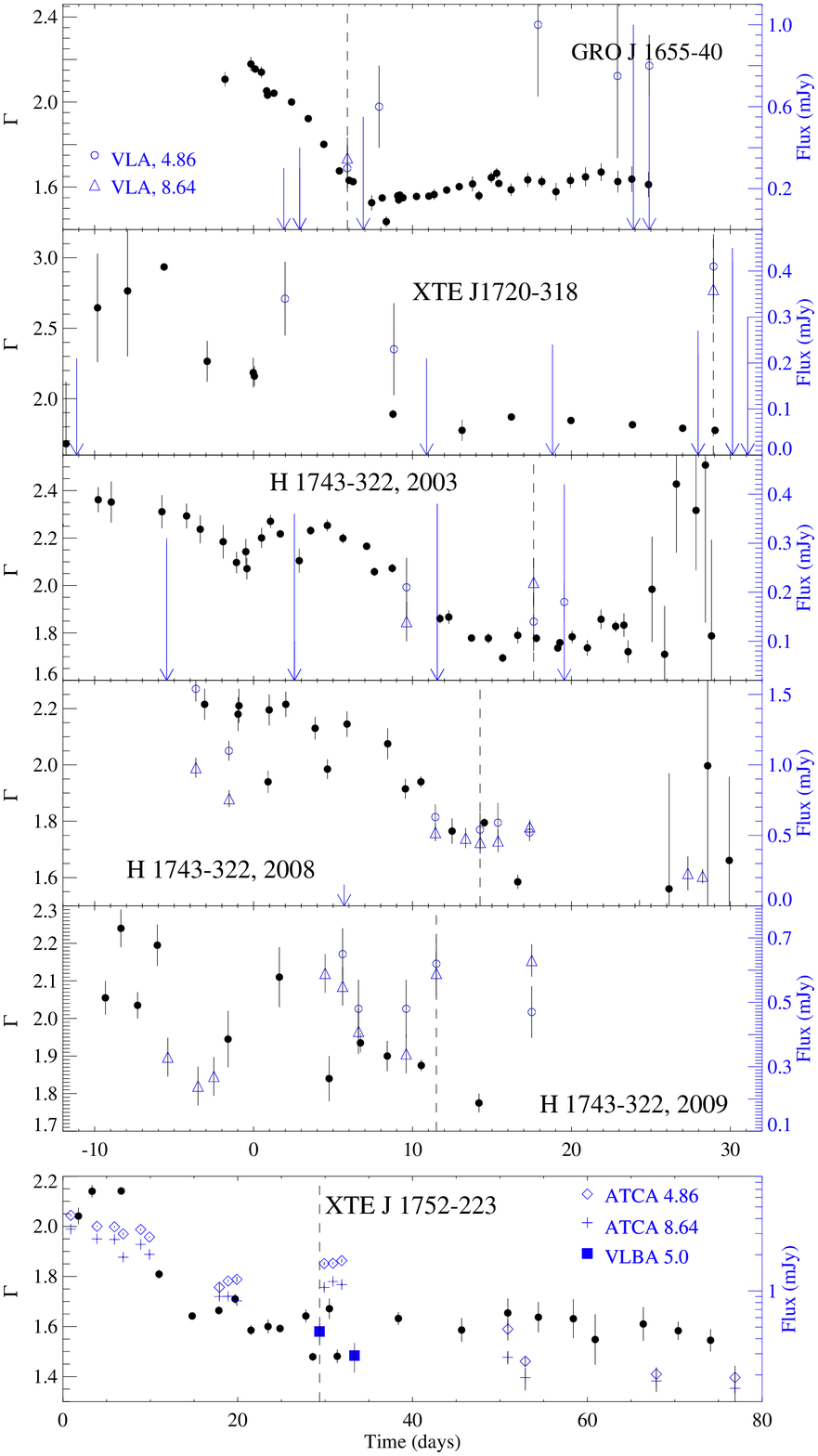}
\caption{\label{fig:harden2}
Evolution of power law photon indices ($\Gamma$, black solid circles)  along with evolution in radio (see legend). The dashed lines show the time beyond which the jet is optically thick. 
 A color version of this figure is available in the online version.
}
\end{figure}

We also investigate the evolution of radio flux and spectrum to find the times that the emission becomes optically thick, indicating compact, steady jets (see Figure~\ref{fig:harden2} for the evolution of radio fluxes and X-ray photon index for sources not shown in Fig.~\ref{fig:harden1}). For observations with multi-frequency spectra, we define the radio transition as the time when the radio spectral indices of the particular observation and the remaining observations are greater or equal to zero within the 1 $\sigma$ error. These dates are shown with dashed lines in Figure~\ref{fig:harden2}, and tabulated in Table~\ref{tabtr} (labeled "Compact") along with the dates of the first radio detections of the sources (labeled "First"). Details of observations for individual sources are provided in the Appendix.

\subsection{The sources}\label{subsec:sources}

For the systematic analysis, we use 7 sources in 12 outburst decays.  For general information on most of these sources and outbursts, see \cite{Dunn10}. The black hole mass and distance estimates that we used are summarized in Table~\ref{table:md}. Specific information is given below:

\GX:  The \rxte\ data from this recurrent source have been analyzed extensively in all outbursts. We utilized four outburst decays:  2003 \citep[MJD 52,680$-$MJD 52,780,][]{Kalemci06_mqw}; 2005 (MJD 53,459$-$MJD 53,496); 2007 \citep[MJD 54,220$-$MJD 54,265,][]{Kalemci08_kol}, and 2010 \citep[MJD 55,560$-$MJD 55,650,][]{Dincer12}. The data and details of the analysis of the SMARTS observations can be found in \cite{Buxton12}. The ATCA radio fluxes are taken from \cite{Corbel12, Corbel13}. 

\FU: We utilize SMARTS for NIR, ATCA and MOST for radio, and \rxte\ for X-rays for the decay of the 2002 outburst of the source, between MJD 52,464 and MJD 52,499 \citep{Kalemci05, Buxton04}. 

\SF: We utilize the VLA\footnote{\url{http://www.aoc.nrao.edu/~mrupen/XRT/GRJ1655-40/grj1655-40.shtml}} radio and \rxte\ observations of this source during the 2005 outburst decay, between MJD 53,625 and MJD 53,645. The evolution of X-ray spectral parameters can be found in \cite{Kalemci06_mqw}.  The SMARTS light curve of \SF\  for the 2005 outburst is dominated by the nearly periodic emission coming from the companion, which is a bright F-type subgiant \citep{Foellmi06}, and there is no secondary flare \citep[see][for the J-band light curve of this source]{Dincer08}.


\FF: We use the \rxte\ and SMARTS data from the 2001 decay of this source, between MJD 51,669 and MJD 51,703. The details of the \rxte\ analysis can be found in \cite{Kalemci01} and \cite{Tomsick01}. The SMARTS data are obtained from \cite{Jain01_b}. See also \cite{Russell10} for a detailed multi-wavelength analysis of the decay of 2001 outburst.

\ST: \rxte\ and SMARTS observations of this source during the decay of the 2010 outburst (MJD 55,240$-$MJD 55,370) have been analyzed by \cite{Chun13}. See also \cite{Russell12} and \cite{Ratti12} for an in-depth discussion of multiwavelength observations during the decay of the outburst. The radio data are taken from \cite{Brocksopp13} and \cite{Yang11}. 

\HS: This source has no NIR coverage due to its position in the Galactic plane but is covered amply in radio during its outbursts. We include data from three outburst decays for this source. For the 2003 outburst decay, the X-ray and radio data are from \cite{Kalemci06} and \cite{McClintock09}, respectively. For the 2008 and 2009 outburst decays, we conducted X-ray spectral and timing analysis as described in \S\ref{subsec:rxtespec} and \S\ref{subsec:rxtetim}.The hydrogen column density is fixed to 2.3 $\times$ 10$^{22}$ cm$^{-2}$ following \cite{Kalemci06}. For radio fluxes, we used \cite{Jonker10} for the 2008 outburst, and \cite{MillerJ12} for the 2009 outburst.


\STT: This source has moderate radio coverage at the rise of the outburst; however, the coverage is not as good during the decay. For the radio we used VLA and ATCA data taken from \cite{Brocksopp05}. It is also followed in NIR and optical, but there are only a couple of observations during the decay \citep{Chaty06}. We conduct spectral and timing analysis as described in \S\ref{subsec:rxtespec} and \S\ref{subsec:rxtetim}. The hydrogen column density is fixed to 1.2 $\times$ 10$^{22}$ cm$^{-2}$ following \cite{CadolleBel04}.



\epsscale{1.25}
\epsscale{1.0}

\begin{table*}
\centering
\caption{Transition times and Eddington Luminosity Fractions\label{tabtr}}
\begin{tabular}{l|cc|cc|cc|ccc}
\hline \hline
 & \multicolumn{2}{c}{Timing Transition (TT)} & \multicolumn{2}{c}{Index Transition (IT)} & \multicolumn{2}{c}{NIR Transition} & \multicolumn{3}{c}{Radio Transition} \\
Source, Year  &  Date & ELF & Lag\footnote{All lags are with respect to the timing transition} & ELF & Lag & ELF & First\footnote{Time of the first radio detection with respect to the TT} & Compact\footnote{Time of the first flat/inverted radio spectrum, or the first detection of the compact core with respect to the TT } & ELF\footnote{ELF of the Compact radio transition} \\
 & (MJD) & (\%) & (days) & (\%) & (days) & (\%) & (days) & (days) & (\%) \\
\hline
GX339-4, 2003 & 52717.8$\pm$0.3 & 0.88$\pm$0.61 &  6.6$\pm$0.3 & 1.34$\pm$0.55 & 22.2$\pm$2.8 & 1.70$\pm$1.17 & $-$ & $-$ & $-$\\
GX339-4, 2005 & 53461.3$\pm$1.8 & 0.90$\pm$0.62 &  5.4$\pm$0.7 & 1.89$\pm$0.41 & 15.1$\pm$1.0 & 2.49$\pm$1.71 & $\textless\,$20.4 & $\textless\,$20.4 & 2.85$\pm$1.30\\
GX339-4, 2007 & 54228.0$\pm$0.4 & 1.22$\pm$0.84 &  5.0$\pm$0.3 & 1.21$\pm$0.76 & 12.9$\pm$1.3 & 2.28$\pm$1.57 & $\textless\,$23.7 & $\textless\,$23.7 & 1.54$\pm$0.83\\
GX339-4, 2011 & 55594.0$\pm$0.7 & 1.05$\pm$0.72 &  2.2$\pm$0.5 & 1.40$\pm$0.67 & 12.2$\pm$1.1 & 1.43$\pm$0.98 &  4.9 & 16.1 & 1.24$\pm$0.96\\
4U1543-47, 2002 & 52473.7$\pm$0.5 & 1.54$\pm$0.53 &  5.7$\pm$0.4 & 1.18$\pm$0.34 &  9.9$\pm$1.0 & 1.31$\pm$0.45 & $\textless\,$13.3 & $\textless\,$16.3 & 0.23$\pm$0.41\\
XTEJ1550-564, 2000 & 51674.0$\pm$0.6 & 3.05$\pm$0.89 &  0.6$\pm$0.4 & 3.05$\pm$0.89 & 12.3$\pm$1.0 & 1.70$\pm$0.50 & $-$ & $-$ & $-$\\
XTEJ1752-223, 2009 & 55282.1$\pm$2.1 & 0.91$\pm$0.60 &  3.7$\pm$0.9 & 1.89$\pm$0.48 & $-$ & $-$ & $\textless\,$ 0.0 & $\textless\,$29.4 & 1.97$\pm$1.24\\
GRO J1655-40, 2005 & 53628.1$\pm$0.1 & 0.75$\pm$0.15 &  1.3$\pm$1.1 & 1.17$\pm$0.13 & $-$ & $-$ & $\textless\,$ 5.9 & $\textless\,$ 5.9 & 1.21$\pm$0.24\\
H1743-322, 2003 & 52930.4$\pm$0.5 & 1.73$\pm$0.65 &  5.7$\pm$0.2 & 1.69$\pm$0.50 & $-$ & $-$ & $\textless\,$ 9.6 & $\textless\,$17.6 & 0.66$\pm$0.64\\
H1743-322, 2008 & 54488.3$\pm$0.9 & 2.45$\pm$0.92 &  8.8$\pm$0.6 & 2.05$\pm$0.76 & $-$ & $-$ & $\textless\,$11.4 & $\textless\,$14.3 & 1.81$\pm$0.77\\
H1743-322,     2009 & 55014.7$\pm$1.6 & 2.35$\pm$0.88 &  9.0$\pm$1.0 & 1.60$\pm$0.80 & $-$ & $-$ & $\textless\,$ 0.0 & $\textless\,$11.5 & 1.26$\pm$0.60\\
XTE J1720-318, 2003 & 52726.6$\pm$0.0 & 0.56$\pm$0.38 & -5.0$\pm$1.0 & 0.30$\pm$0.41 & $-$ & $-$ & $\textless\,$ 2.0 & $\textless\,$28.9 & 0.97$\pm$0.20\\
\end{tabular}
\end{table*}




\section{Results}\label{sec:results}

In our prior work, we showed that there are certain changes in X-ray spectral and timing properties of GBHTs along with changes in the NIR and radio properties \citep{Kalemci06_mqw, Kalemci08_kol}. We established that during outburst decay, a sharp change in the timing properties takes place first, with an abrupt increase in the rms amplitude of variability accompanied with an increase in the power-law flux \citep{Kalemci03}. This transition is called the ``timing transition" (TT) in this work (see the Appendix for the details of how we determine the TT for each source). For figures starting from Fig.~\ref{fig:allflux}, the observations before this transition are shown with orange in the color version and are denoted as ``Before TT". This transition is also the reference date that is denoted with time 0 in the figures. Note that the TT is often not associated with a change in the power-law index.  

The next important change is the significant hardening of the spectra (see Figs.~\ref{fig:harden1}, \ref{fig:harden2}). This transition is called the index transition (IT) in this work (see Appendix for the details of how we determine the IT). In the figures, the observations after the TT but before the IT are shown with green in the color version and denoted as ``After TT / before IT".  The next transition is the increase in the NIR flux and/or radio detection of the compact jet. Since we assumed that the NIR increase is due to the formation of the compact jet, we define a ``compact jet transition" (CJT). For \GX, \FU, and \FF\ the CJT corresponds to the "NIR Transition", and for the rest of the outbursts, it corresponds to the "Compact" radio transition in Table~\ref{tabtr}. The observations before CJT and after the IT are shown with blue in the color version and denoted as ``After IT / before CJT". Finally, all observations after the CJT are shown in red in the color version.


\subsection{Transition luminosities}

\begin{figure*}[t]
\plotone{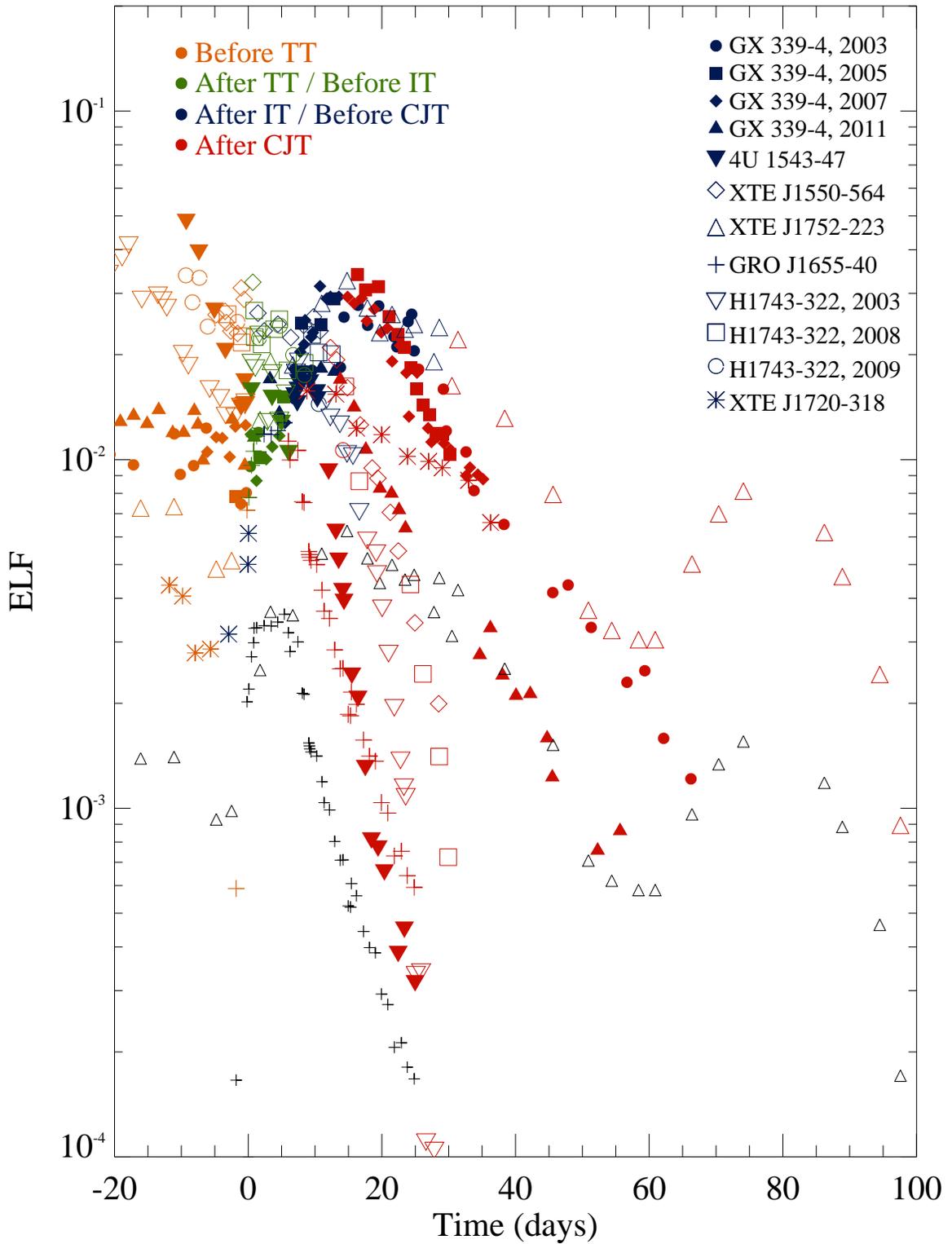}
\caption{\label{fig:allflux}
The evolution of 3-200 keV Eddington Luminosity Fraction (ELF) of all sources investigated in this work.  The distances and black hole masses that are used to calculate luminosities are shown in Table~\ref{table:md}. For \ST, and \SF,  the luminosities are calculated twice for this graph only; the smaller black points at lower Eddington Luminosities are from  distance measurements of \cite{Shaposhnikov10} for \ST, and \cite{Foellmi06} for \SF. The errors in the luminosities are large for some systems due to large errors in distance. The errors are not shown for clarity, but incorporated in the measurements and the discussion. A color version of this figure is available in the online version.
}
\end{figure*}

In Fig.~\ref{fig:allflux}, we plot the evolution of the Eddington Luminosity Fraction, ELF, of all sources we investigated as a function of time. The color scheme is explained above. As already established by \cite{Maccarone03_b}, the transition to the hard state occurs at similar ELFs. This figure includes two sets of data for \ST\ and \SF\ due to differences in distance measurements by different groups. The data with distance values given in Table~\ref{table:md} are shown with large symbols, and the alternative cases are shown with black, smaller symbols. For later figures,  we do not show data with alternative distances since they can easily be scaled. Also, the plots that used ELFs do not include error bars. Due to large uncertainties in the distance measurements, the errors in luminosities are large (up to 50\% for \GX\ and \ST) and clutter the figures. Thus, the figures with luminosities should be regarded with some caution. The errors are incorporated in the measurements and discussion. The ELFs and times of transitions with respective errors can be found in Table~\ref{tabtr}.

\epsscale{1.1}
\epsscale{1.0}

\epsscale{1.15}
\begin{figure*}[t]
\plottwo{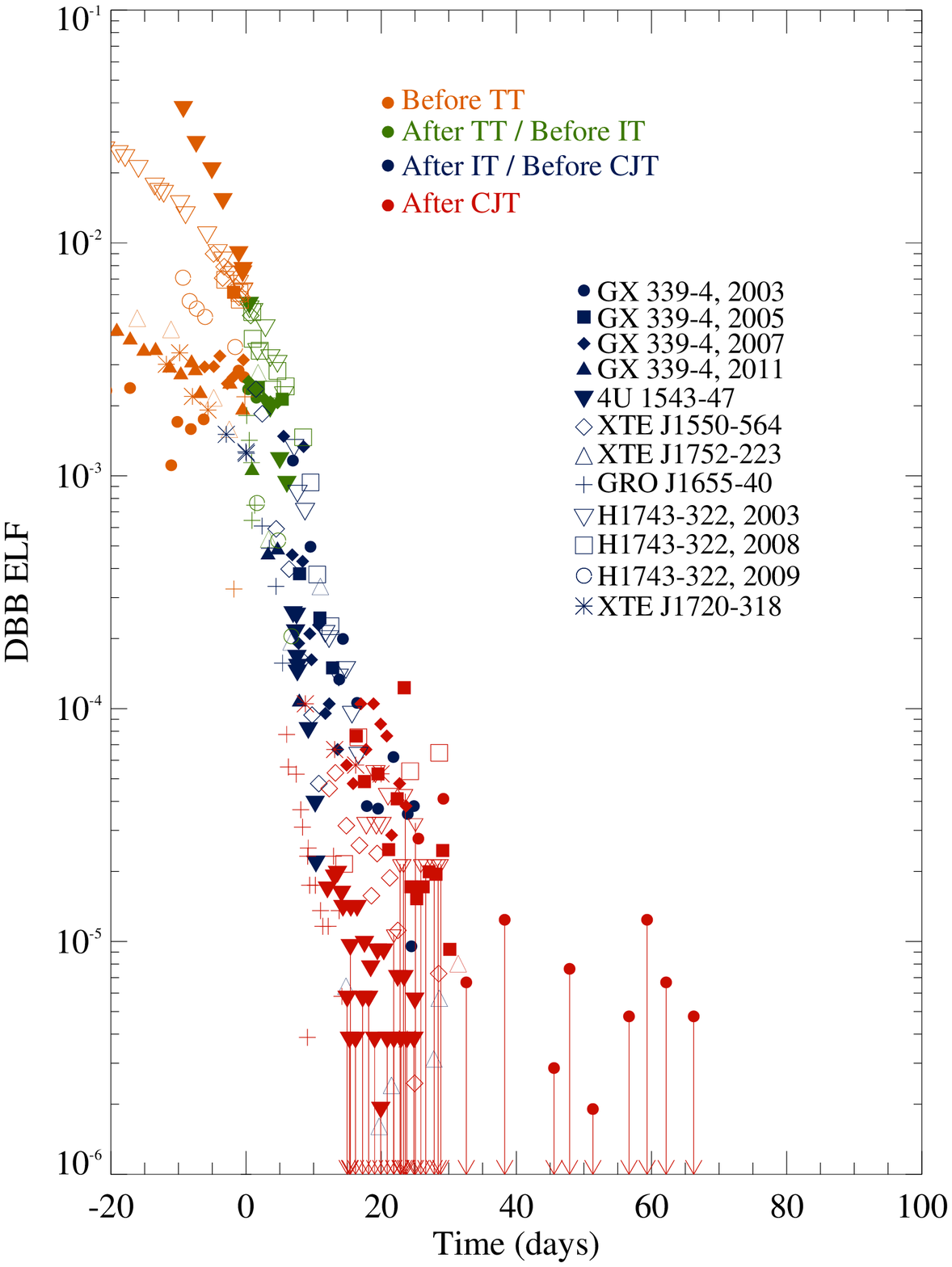}{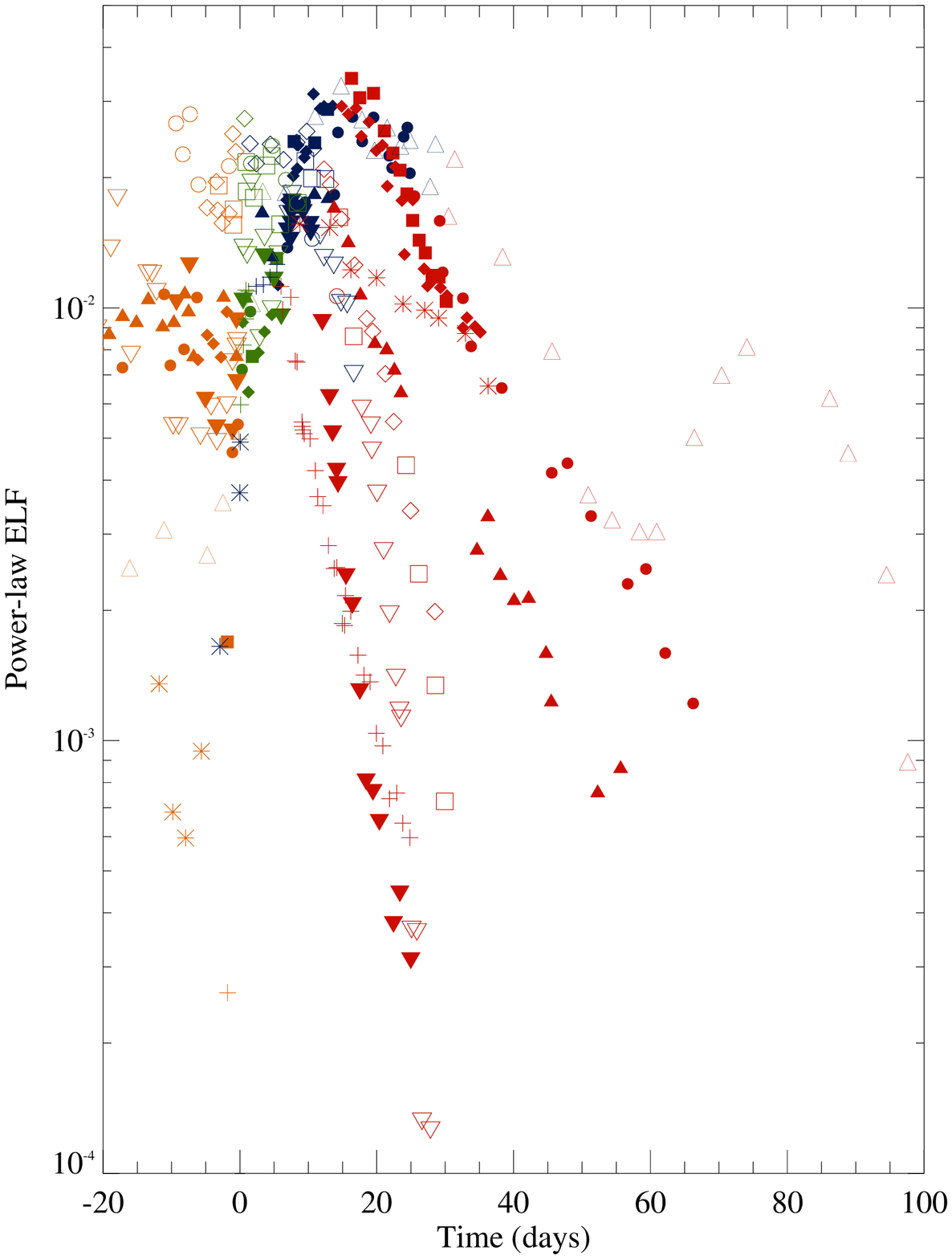}
\caption{\label{fig:dbbflux}
The evolution of 3-200 keV disk blackbody (left), and power-law (right) ELF of all sources investigated in this work. 
A color version of this figure is available in the online version.
}
\end{figure*}
\epsscale{1.0}

The disk blackbody (diskbb) ELF evolution provides striking patterns during important spectral changes during the decay (see Fig.~\ref{fig:dbbflux}, left). The decays of the diskbb ELF are exponential for each source, but the decay rate is different before and after the TT. The diskbb ELF decays faster after the TT. More importantly, there is a definite threshold, the diskbb ELF must be below \wsim 0.0001 for the compact jets to form during the outburst decays.


The TT is often associated with an increase in the power-law flux as it can be seen in Fig.~\ref{fig:dbbflux}, right. The power-law ELFs are more scattered compared to those of diskbb ELF. The most important point about the evolution of power-law ELFs is the fact that the compact jets almost always form after the power-law ELF peaks. The lag between the peaks of power-law ELF and compact jet formation is often days.

\epsscale{1.2}
\begin{figure}[t]
\plotone{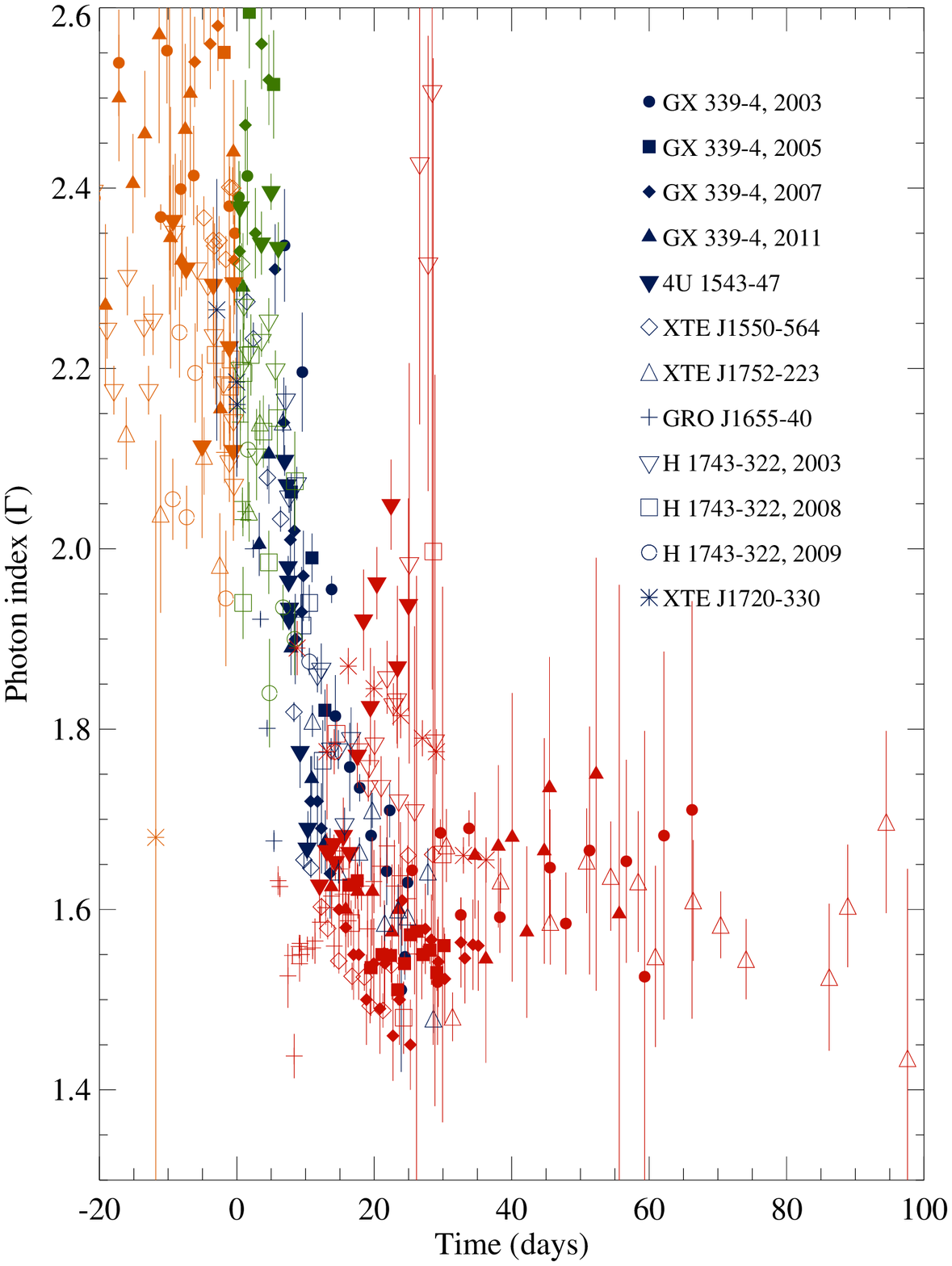}
\caption{\label{fig:ind}
The evolution of power law photon indices ($\Gamma$) for all sources investigated in this work. A color version of this figure is available in the online version.}
\end{figure}
\epsscale{1.0}

The transition from an intermediate state to the hard state is evident in all sources as a fast hardening of the photon index as shown in Fig.~\ref{fig:ind} blue points in the color version. Excluding \STT\ and \ST, there is a threshold also in the power-law index of around 1.8. This result is spectral model dependent, and rather than providing a quantitative threshold, a better statement would be that the compact jets form when the source is close to being at its hardest in the X-ray band during outburst decay. Moreover, this plot also shows the clear softening of some of the sources at the end of outbursts as discussed before by \cite{Tomsick01, Corbel06, Dincer08, Wu08, Sobolewska11}.


\section{Discussion}\label{sec:discussion}

\subsection{NIR flares, radio detections and jet formation}\label{subsec:rebright}

The jet, which is quenched in the soft state, turns on during outburst decay. The X-ray spectral and temporal changes may allow us to understand the necessary environment and timescale for the jets to be launched and to evolve. We have very good coverage in the NIR; however, our radio coverage is sparse. The SEDs prepared during the flares for \FU\ and \FF\ are consistent with emission from compact jets \citep{Kalemci05, Russell10}. There is a radio detection of \FU\ with an inverted radio spectrum during the flare. The SEDs prepared from the data of \GX\ during the flares are harder to interpret because the NIR-optical part of the SEDs are rather flat \citep{Buxton12,Dincer13} which require extra emission components in the jet. For the 2005 and 2007 outburst decays of \GX, a compact jet is detected close to the peak of the NIR flare. In the case of the best radio and NIR coverage, \GX\ in 2011, we observe that there are radio detections earlier than the NIR peak, with an optically thin spectrum, and the radio spectrum becomes optically thick after the NIR flare starts. There is no radio observation during the flare of \FF, but given its SED, the morphology of the NIR flare compared to those of \GX\ and \FU, and the time it starts compared to the timing transition (see Fig.~\ref{fig:harden1})) it is reasonable to assume that NIR flares of \FU, \GX\ and \FF\ all have the same origin, and they are all related to compact jets. 

For \GX, \FF\ and \FU, the delay between the TT and the start of the NIR peak is 10-20 days (see Fig.~\ref{fig:harden1}). For those sources, the NIR peak occurs 5-15 days after the index transition.  When we investigate the decays with radio coverage, there are cases with radio detections earlier than even the timing transition. However, the first radio detection does not always mean the presence of a compact jet with a flat/inverted radio spectrum (see Table~\ref{tabtr}). The first two radio observations of \STT\ are taken at a single frequency, therefore we do not know if it is optically thin or thick. The radio spectra of \HS\  for all outburst decays evolve from optically thin to optically thick (see Fig.~\ref{fig:harden2}).  

 In fact, with the strict definition of presence of flat to inverted radio spectrum for the radio transition, all sources with both NIR and radio coverage show compact jet ``after" the NIR flare start. The radio behavior of all sources is consistent with what is observed in \GX\ in 2011, as the X-ray spectrum gets harder, a detection or increase in radio flux is observed first. When the source is close to its hardest level, the NIR flux rises (for the cases with NIR coverage), and the radio spectrum becomes flat or inverted \citep{Corbel13}.  These results indicate that if the NIR flare has a jet origin, it corresponds to a change in properties of the jet (becoming compact, and therefore, optically thick to its own radio emission) rather than indicating the time of the jet launch.  A similar explanation is also given by \citet{MillerJ12}.

 The optically thin radio emission may be coming from a jet that was launched earlier interacting with the interstellar medium, or there may be an outflow which is not collimated enough to produce a flat to inverted radio spectrum. 
 
The case of \ST\ warrants a separate discussion. As shown in \cite{Chun13}, the $I$ and $H$ band SMARTS light curves indicate three possible flares.  When the compact core is detected with the VLBI ($\wsim$ 29 days after the TT), a small flare (flare 2 in \citealt{Chun13}) in the $I$ band is in progress. We note that the ATCA radio spectrum is still optically thin at this time. The ATCA radio spectrum becomes consistent with emission from a compact jet during a larger flare (flare 3 in \citealt{Chun13}) observed both in the $I$ and the $H$ bands $\wsim$ 50 days after the TT \citep{Chun13}. If flare 2 in the $I$ band is similar to flares seen in \GX\ and \FU, then the scenario discussed above is also valid for \ST. Only for the case that flare 3 is similar to flares seen in \GX\ and \FU, and flare 2 is due to some other process, we must conclude that the optically thick jet is launched tens of days before the NIR flare, and therefore the scenario that the NIR flare corresponds to the transition of optically thin to optically thick radio emission is not valid for all sources.





\epsscale{1.2}
\begin{figure}[t]
\plotone{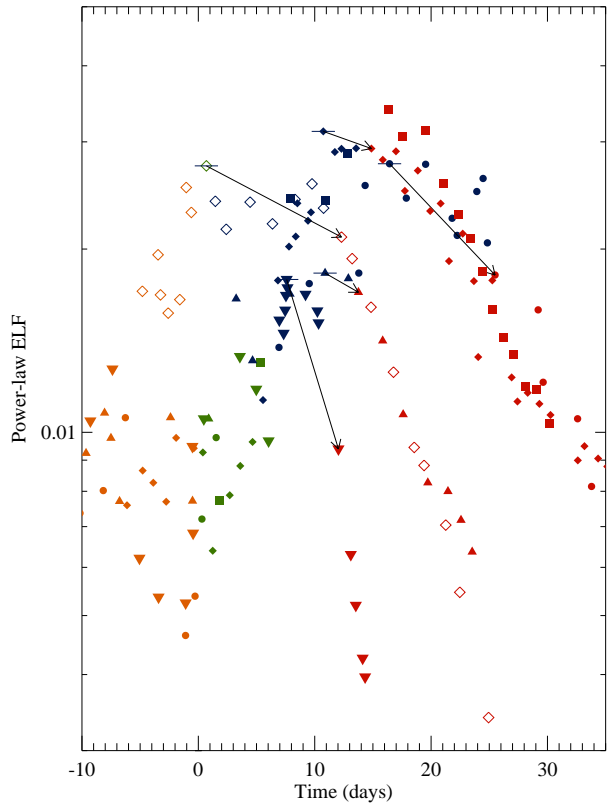}
\caption{\label{fig:zoom}
Zoomed region of the evolution of power-law ELF shown in Fig.~\ref{fig:dbbflux}, right. The arrows are from the peak ELF to the first observation after CJT to show the timescales. Only sources with well determined transition times are shown. A color version of this figure is available in the online version.}
\end{figure}
\epsscale{1.0}

\subsection{Timescales of transitions}

The time between the TT (sudden increase in the rms amplitude of variability)  and the IT (start of the hardening) is usually between 3-7 days. It is well known that the increase in the variability rms is linked to the presence and strength of a hot electron corona \citep[even though the origin of variability pattern may still be the optically thin disk,][] {Uttley11}. The increase in the power-law flux during this transition supports the connection between the corona and timing (see Fig.~\ref{fig:dbbflux}, right). This increase in flux from the corona does not immediately result in the change in the photon-index (see Fig.~\ref{fig:ind}, \citealt{Kalemci03}). 

In our previous work, we claimed that, to form the jet, a strong corona must be formed first \citep{Kalemci05, Kalemci06} based on limited data and theoretical arguments \citep{Meier01, Meier01_b}. Here, we try to determine the timescale between the establishment of the strong corona and the CJT. We chose the date at which the power-law ELF peaks as a safe upper limit for the presence of a strong corona. The delay timescales between the times for which the power-law ELF peaks and the CJT are between 0-12 days (see  Fig.~\ref{fig:zoom}).  As stated in \cite{Fragile12}, three ingredients are required to launch jets: accretion of large scale poloidal magnetic fields, collimation, and mass loading/energy conversion to electromagnetic emission.  The simulations in general also show that once the poloidal magnetic field is transferred, the outflows are seen on a dynamical timescale \citep[see][and references therein]{Meier12}, which is obviously multiple orders of magnitude shorter than the delay that we determined for most of the sources in Fig.~\ref{fig:zoom}. 

\cite{Corbel13} discuss the delay between the optically thin radio detection (which occurs after the IT, but before the power-law ELF peaks) and the NIR rise in \GX\ in terms of a timescale for building up turbulence and strong shocks in the magnetized jet plasma. Initially, the plasma density and therefore the optical depth is low, producing optically thin radio emission. In this scenario, an increase in the optical depth and/or an increase in particle acceleration efficiency lead to NIR emission from the jet base turning on.

 According to the theoretical work of \cite{Falcke04} and \cite{Heinz03}, the jet power is positively correlated with the spectral break between the flat and the optically thin part of the jet SED. As the jet power increases, the spectral break shifts to higher frequencies, allowing for the detection of the NIR flux \citep{Coriat09, MillerJ12, Russell13b}. In this case, the delay of the NIR peak with respect to the other changes in X-ray properties depends on the evolution timescale of the SED break. Such an evolution of the break frequency is observed for MAXI~J1836$-$194 during the decay of a failed outburst \citep{Russell13a}. However, in that case, a compact jet with a flat to inverted spectrum never disappeared (since the source never entered the soft state) and is present even before the NIR rise. It is difficult to test such a scenario with the data we have as none of the three \GX\ outburst decays with simultaneous NIR and radio data show flat to inverted radio spectrum before the NIR flare.

If the increase in the NIR flux is due to the formation of a compact jet, then the delay between the transition in timing and the compact jet formation may also be due to the transport of poloidal magnetic field from a very large distance in viscous time scales. The poloidal magnetic field would increase particle acceleration efficiency. It is also claimed that it is easier to transport magnetic fields if the disk is geometrically thick \citep{Beckwith09}.  On the other hand, there are also reports claiming that the thickness of the disk is not an important factor in the transport of magnetic fields and that thin disks would carry them as efficiently \citep{McKinney12}. 

This raises another question: what is the reservoir of the poloidal fields that launch the jet? The fields can be generated through dynamo mechanisms at the outer parts of the disk, but it is also possible that the source is the secondary star \citep{McKinney12}. Then, the transport timescale could still be a plausible explanation if the efficiency of transport is accelerated by irradiation of the secondary star by hard X-rays after the TT.

\subsection{The optically thick, geometrically thin disk}

The answer to the puzzle of what sets the timescale for the compact jet production may be found in the evolution of the optically thick disk. Fig.~\ref{fig:dbbflux} clearly indicates a threshold for the CJT at $10^{-4}$ ELF. This threshold was first reported in \cite{Kalemci06} based on only a couple of sources.  With the new additions of sources and outbursts, it has become much clearer. To be able to observe the compact jet, the disk flux in the \rxte\ band must be lower than a threshold value while the corona has already formed as seen by the evolution of the power-law flux. 

The disk luminosity may be affecting the disk height, and therefore the collimation of the jet directly.  However, it is shown by \cite{Fragile12} that the disk scale height is not an important factor in the collimation of the jet, nor does it affect the jet power. In their simulations, the collimation is provided by the corona. Alternatively, the disk height could affect the transport properties of the magnetic field, but as discussed earlier, there is no consensus on the role of disk thickness in the efficiency of magnetic field transport.

 The launch of compact jet may require a truncated disk. There is evidence of disk truncation at very low ELF for \GX\ \citep[0.0014 ELF for total flux,][]{Tomsick09}. However, there is evidence of the disk being close to the last stable orbit when the jets are present for \GX\ \citep{Tomsick08, Allured13}. A recent, detailed study by \cite{Miller12} shows not only that the jets are present with the disk being close to the last stable orbit, but also that the jet flux is not related to the inner disk radius for Cyg X-1. We note that the inner disk structure could be formed by the coronal condensation, and at those luminosity levels the accretion can still be dominated by a coronal flow \citep{MeyerHof12}.

If we put these pieces of information together, a picture emerges that can explain the relation between the NIR evolution, radio emission and jets.  X-ray spectral evidence in many black hole transients indicate that the corona is very small and/or patchy in the soft state,  and the jet emission is quenched. After the TT, there is an increase in power law flux, indicating formation of a stronger corona. Weak, optically thin outflows may form at this time as observed in some sources, and as predicted by simulations. On the other hand, in this work, we show that even when the coronal emission totally dominates the accretion flow, as indicated by the maximum power law ELF, the compact jets may not form. As the disk flux decreases below a threshold, the corona fills a larger area. Once a large corona is formed, it produces very hard X-ray emission through Comptonization (of either the disk photons or soft photons that are created by synchrotron radiation at the base of the jet \citep{Markoff05}) and, at the same time, very effectively collimates the jet. At this point, the optical depth of the corona is high enough for the radio spectrum to be optically thick, and the jet base efficiently produces NIR photons. The change in the properties of the corona may also aid in transport of magnetic fields.

\subsection{Comparison with the outburst rise}

The fact that there is a threshold diskbb ELF for the formation of the compact jet does not mean that the jets will disappear when the diskbb ELF rises above the threshold. The hysteretic behavior of the state transitions in GBHTs is very well known.  During the rise, the transitions take place at higher luminosities than during the decay.  As expected, the compact jets observed during the outburst rise persist within much larger disk blackbody ELFs \citep{Joinet05, CadolleBel11}. As far as we know, there has not been a systematic study of the relation of the disk flux to the NIR behavior during the outburst rise. However, \cite{Homan05} investigated the 2005 outburst rise of \GX. The NIR light curve shows a flare in the initial hard state that rolls off slightly around MJD~52,400, and then sharply drops off 4 days later. This drop-off is accompanied by a rise in the disk flux by almost two orders of magnitude. \cite{CadolleBel11} analyzed multiwavelength data from the rise of the 2010 outburst of \GX. The disk behavior is similar to 2005: The $i$, $V$ and $R$ band magnitudes show a sharp decrease on MJD~55,294. The optical flux continued to drop over 6 days while the disk flux rises sharply. It would not be surprising if the large increase in the disk flux changes the magnetic field structure of the environment and quenches the compact jets. The observations during the rise also show the importance of the emission from the optically thick disk in terms of determining the behavior of the compact jet. There is, though, a major difference between the rise and decay behavior of the disk in addition to the different ELF thresholds. For \GX\ during the 2010 rise, the jet persists when the ratio of the disk flux to overall flux is around 20\%. When we place the observations with the compact jet in the hardness-intensity diagrams \citep{Fender04}, it can be clearly seen that the compact jets persist to much lower hardness levels in the rise compared to the decay \citep{Fender09,Dunn10}. Therefore, forming and sustaining compact jets have different requirements in terms of X-ray emission from the disk.

\subsection{Alternative explanations for the NIR peak}

Synchrotron emission from a hot accretion flow \citep{Veledina11a, Veledina12} is a viable alternative to the scenario we have been considering. In this work, we discuss the case of synchrotron emission from an optically thick compact jet for the source of NIR photons during the peak.  A hot accretion flow model assumes that synchrotron emission from the corona is caused mostly by non-thermal electrons. The flat NIR-optical SEDs observed for GX~339$-$4 \citep{Dincer12, Gandhi11} are consistent with the hot accretion flow scenario. The coincidence of optically thin to optically thick radio transition with the NIR increase in 2011 can also be explained within this model as the presence of a large corona could explain both the enhanced synchrotron emission from the hot accretion flow, and enhanced collimation of the jet. However, we show in this work that the CJT occurs as much as 12 days after the power-law ELF peaks. If the strongest power-law ELF is an indication of a large corona filling the inner parts of the accretion disk, the delay of the NIR peak is difficult to understand in the hot accretion flow model.  Moreover, this model cannot explain the SEDs with steep slopes in the NIR-optical region after the break seen in \FU\ \citep{Kalemci05} and MAXI~J1836$-$194 \citep{Russell13a}. 

Finally, irradiation related emission from the outer parts of the disk or from the secondary star should be considered for the secondary NIR rise. According to \cite{Homan05}, the contribution of the secondary star is negligible for \GX\ during outbursts. It is also shown in \cite{Buxton12} and \cite{Dincer12} that the optical-infrared part of the SED of \GX\ is consistent with viscously heated disk emission in the soft state, but as the NIR flux rises it becomes redder. \cite{Coriat09} claims that the delay between the transition to the hard state and the rise of the NIR rules out the irradiation scenario. However, geometrical effects such as the location of the cooling front and the height of the corona must also be taken into account before reaching strong conclusions. As shown in \cite{Ertan02}, the parts of the accretion disk that remain inside the cooling front can become geometrically thicker due to heating from the central source and efficient dissipation with the hot-state viscosity. As a result, during the early, softer part of the decay, the outer disk beyond the cooling front can remain shielded from the X-rays by the hot inner disk. As the source moves towards the hard state, the corona becomes larger while also growing in scale height with respect to the disk. After the corona reaches a critical height, X-rays start irradiating the outermost part of the initially shaded cold disk.  This would be well after the transition in timing (during which the power-law flux increases). This scenario could explain the timing of the flares, however, eventually the newly irradiated regions of the disk will accrete and increase the soft X-ray flux. Among the sample we investigated, only \ST\ showed a secondary brightening in X-rays. The irradiation+cooling flow scenario would not work for  \GX, \FF, and \FU\ flares we study here as there are no associated X-ray peak after the NIR flare. A detailed study is underway to investigate whether this model could explain both the SED evolution and relative timing of optical and X-ray peaks in sources for which both peaks are observed.

We also must note that  if the reservoir of the poloidal magnetic fields is the secondary star, or the outer parts of the accretion disk, the CJT timescales may be explained by an increased efficiency of magnetic field transport due to the irradiation of the outer parts of the disk or the secondary star from a relatively thicker corona (power-law flux increase during the TT). Then, it would take 5-20 days for the ordered magnetic fields to be transported to the inner parts of the disk. This timescale would depend on the size of the disk and the viscosity. If this is the case, the viscosity should vary by a factor of two from outburst to outburst of \GX\ (see Table~\ref{tabtr}). Also, the binary separations of \GX,  \SF, and \FF\ are similar  (Table~\ref{table:md}), but the timescale of CJT for \SF\ is much smaller than those of \GX, and \FF\ (Table~\ref{tabtr}). 

\epsscale{1.2}
\begin{figure}[t]
\plotone{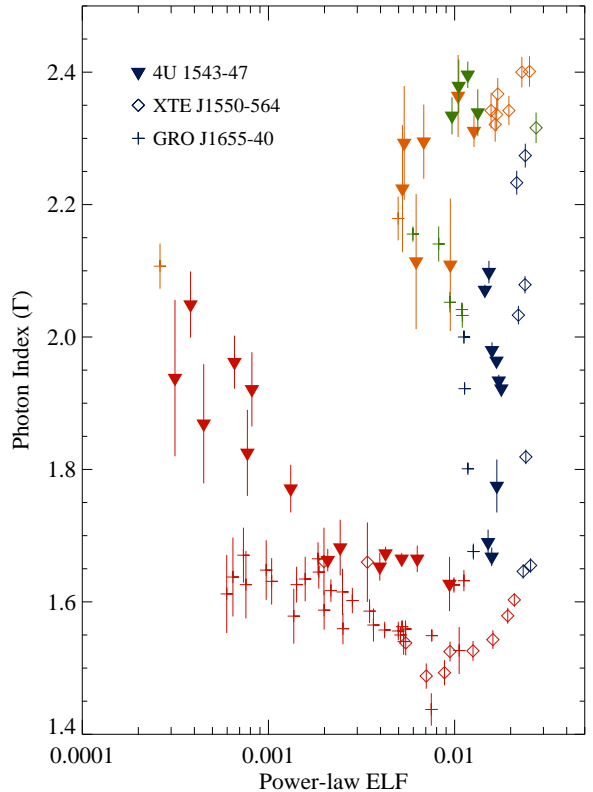}
\caption{\label{fig:plfind}
Photon index ($\Gamma$) vs. power-law ELF for sources that clearly show softening at very low luminosities. The coloring scheme is the same as previous plots. A color version of this figure is available in the online version.
}
\end{figure}
\epsscale{1.0}

\subsection{X-ray softening at the end of outbursts}

Three sources show clear softening at low luminosities during the decay, \FU\ \citep{Kalemci05,Wu08,Dincer08}, \FF\ \citep{Kalemci02,Wu08}, and \SF\ (\citealt{Dincer08}, \citealt{Sobolewska11}, but also see \citealt{Homan13}. On the other hand, our analysis do not show any clear softening for \GX, \ST, or \HS\ in any of their outbursts for the date range investigated in this paper. This is notable because the presence of softening was claimed by \cite{Sobolewska11} for \GX\ in 2003 and 2005 outburst decay. However, they used a Comptonization model rather than a simple power-law to fit the data. Also, their data analysis did not take into account the Galactic ridge emission (M. Sobolewska, private comm.), which could soften the spectrum at low flux levels. We note that fits with a simple power-law model show softening for the 2007 outburst of \GX\  \wsim130 days after the NIR rise \citep{Dincer08}. Concentrating on the cases where there is clear softening, we observe that for \SF, the softening begins almost as soon as the source is detected in radio (see Fig.~\ref{fig:harden2}). \FF\ starts softening at the peak of the NIR rise, and \FU\ softens well after (6-7 days) the NIR peak, as the NIR flux was decaying (see Fig.~\ref{fig:harden1}). They all soften between 0.001-0.01 power-law ELF as seen in Fig.~\ref{fig:plfind}. The softening can be explained both in terms of radiatively inefficient flows or non-thermal jets. For detailed discussions of these alternatives, see \cite{Wu08,Sobolewska11,Russell10}. We would like to emphasize an important finding: the three cases with the clear softening are the ones with the fastest drop in the power-law flux (see Fig.~\ref{fig:dbbflux}, right).  This is consistent with the scenario described in \cite{Russell10}. When the compact jet first forms, the X-rays it produced by synchrotron mechanism may be much less compared to the X-rays produced through thermal Comptonization. Since, for these sources, the hard Comptonization flux drops much quicker, softer X-rays from jet synchrotron may result in a steeper photon index at the end of outburst decays. For \GX, the decay of Comptonized power-law flux is slow, and the effect cannot be seen within 30-40 days after the transition. In fact, for the 2003, 2005 and 2011 outbursts, the NIR flux also decreases within 80 days, but in 2007, the NIR flux stays constant for over 100 days, and the softening in the X-rays is observed as the X-ray flux decreases 130 days after the start of the NIR rise \citep[see][for the evolution of the NIR flux and the power-law flux]{Dincer08}. The drop in X-ray flux is also fast for \HS\ in 2003 and 2008.  However, the Galactic Ridge strongly affects the \rxte\ spectra of this source at low luminosity levels, and the corresponding errors on the photon index make it difficult to detect any indication of softening. 

\section{Summary and Conclusions}

We study the evolution of X-ray spectral properties of all black hole transients during outburst decay for which multiwavelength coverage is also available. We determine the times of several state transitions related to the X-ray and NIR emission: the timing transition (a sudden increase in the rms amplitude of variability, which is accompanied by an increase in power-law flux), the index transition (a slow hardening of the X-ray photon index), and the compact jet transition (a sudden increase in the NIR flux and/or detection of a flat to inverted radio spectrum). The important results of this study are:

\begin{itemize}
  \item State transitions during the decay occur at similar Eddington Luminosity Fractions of a few percent (Fig.~\ref{fig:allflux}; see also \citealt{Maccarone03_b}).
  \item The timescale between transitions is typically between 2-10 days (Table~\ref{tabtr}), which is too long when compared to the jet formation timescales in numerical simulations. The presence of a corona is not enough to launch compact jets (Fig.~\ref{fig:zoom}).
  \item The compact jets are observed when the spectrum is very hard (Fig.~\ref{fig:ind}), but optically thin outflows may start at softer X-ray photon  indices.
  \item There is a strict threshold in the optically thick disk emission, the Eddington Luminosity Fraction of the disk must be smaller than 0.0001 to observe compact jets (Fig.~\ref{fig:dbbflux}).
  \item The sources that show softening at low luminosities are those for which the X-ray flux drops faster. This shows that a secondary, softer component may be affecting the emission for these sources (Fig.~\ref{fig:plfind}).
\end{itemize}

To explain these results we suggest a model that includes a small scale height corona along with the disk, which allows outflows with optically thin emission before compact, steady jets are observed.  In this picture, a large scale height corona develops as the disk emission decreases, providing the collimation for the steady compact jet.  We provide an alternative explanation to the timescale of jet formation in terms of transport of magnetic fields from the outer parts of the disk, and also discuss two alternative explanations for the multiwavelength emission: hot inner accretion flows and irradiation.


\acknowledgments E.K, T.D and Y.Y.C.. acknowledge T\"UB\.ITAK 1001 Project 111T222, the EU FP7 Initial Training Network Black Hole Universe, ITN 215212, and thank all scientists who contributed to the T\"{u}bingen Timing Tools. EK thanks Chris Fragile, David Meier, Rob Fender, \"Unal Ertan, and Juri Poutanen for fruitful discussions. JAT acknowledges partial support from NASA Astrophysics Data Analysis Program grant NNX11AF84G. 



\clearpage

\appendix

\section{Determination of transition times}

\begin{figure}[b]
\plottwo{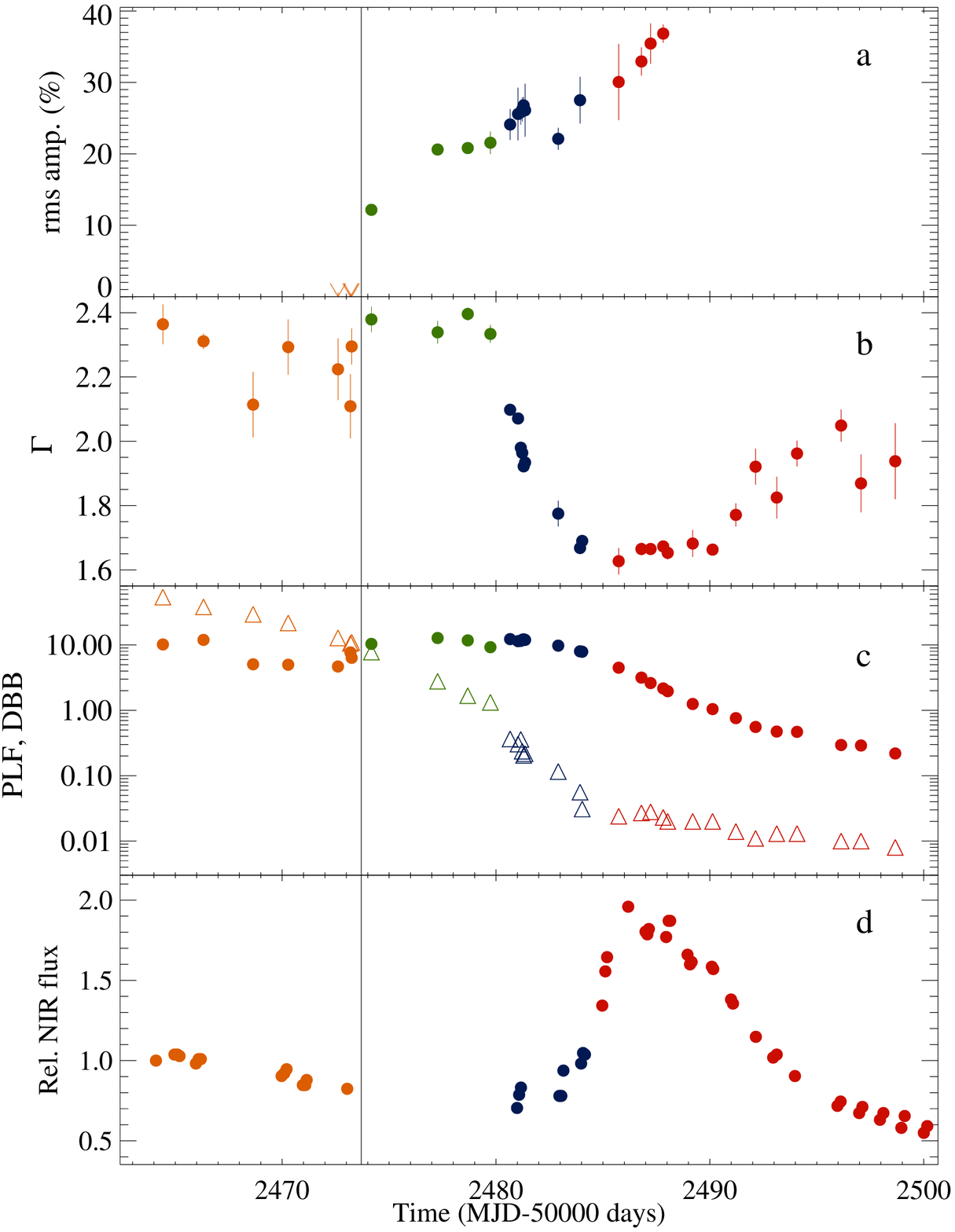}{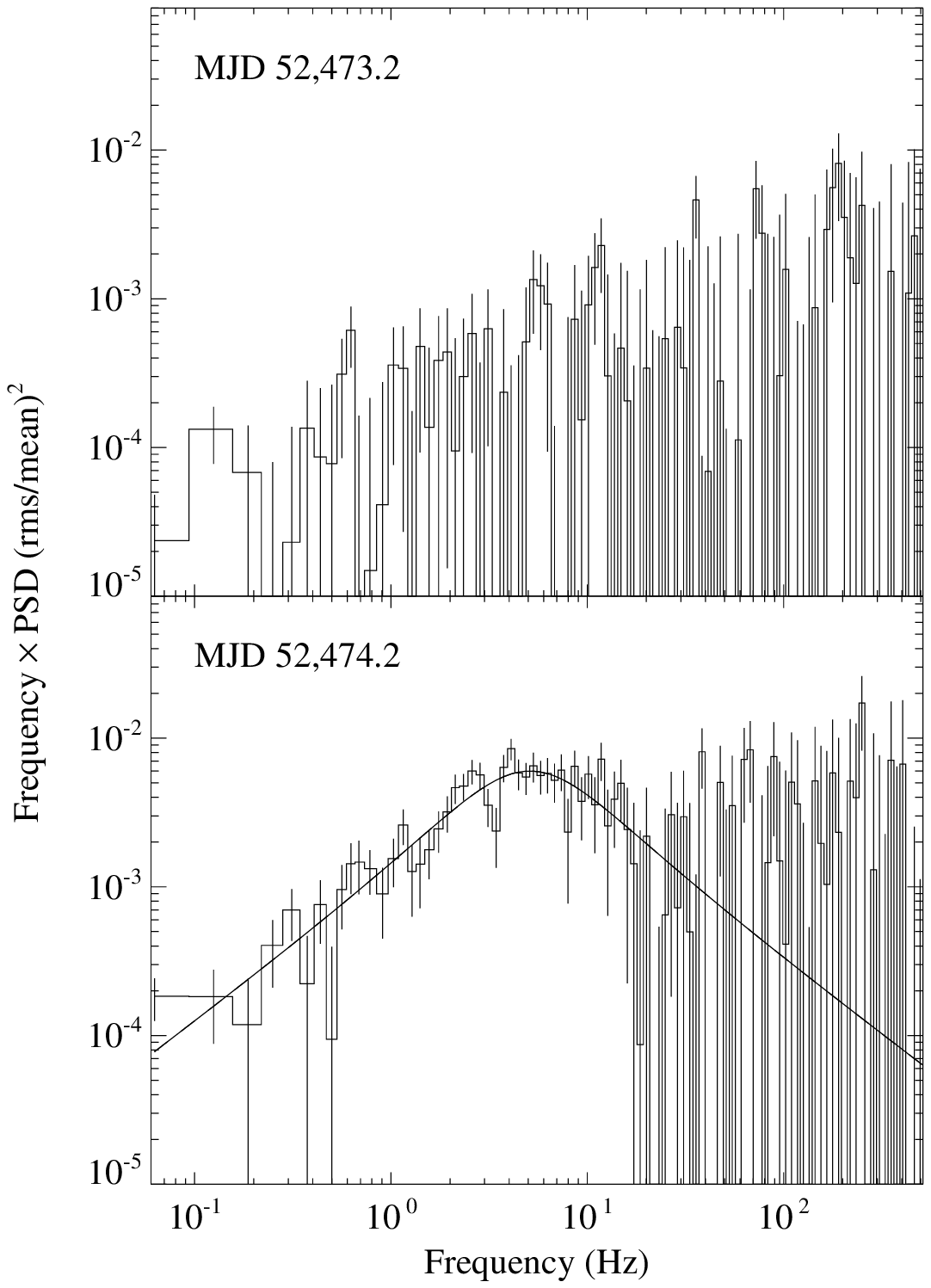}
\caption{\label{fig:fuevol}
Left: Evolution of temporal and spectral parameters of \FU\ during the 2003 outburst decay: a) rms amplitude of variability, b) photon index, c) solid circles power-law flux, triangles diskbb flux in units of 10$^{-10}$ ergs cm$^{-2}$ s$^{-1}$, d) NIR fluxes relative to the first observation. Color definitions are given in \S~\ref{sec:results}. Right: PSD of observations at MJD 52,473.2 (Top) and at 52,474.2 (bottom). The solid line is a fit to the data with a Lorentzian.  
}
\end{figure}

For each source, the specific transition times (timing, index and NIR/radio transition) are obtained as follows. First, we plot the evolution of all spectral and temporal parameters that we investigate during the outburst decay. We provide \FU\ during 2003 outburst case as an example in Figure~\ref{fig:fuevol}.

\epsscale{1.1}
\epsscale{1.0}

\subsection{Timing transition (TT)}

To determine the time of TT, we start with the evolution of PSDs.  For many outbursts (\GX\ in 2005, 2007, 2011, \FU, \SF\ in 2005, \HS\ in 2003, 2008, and \STT), the identification of the TT is straightforward. It occurs when, for two consecutive observations, the PSD in the earlier observation does not show any broad band features with the rms amplitude being less than 5\%, while the later observation shows broad band noise with an rms amplitude of variability greater than 8\% (see the case of \FU, Figure~\ref{fig:fuevol} as an example).  We define the TT to be the temporal midpoint between these two observations.  These sources also show a sudden increase in the power-law flux during this transition as discussed in \cite{Kalemci03}. The detailed evolution of all PSDs for all sources is provided in the following paper \citep{Dincer13}. 

For the rest of the outburst decays we investigate, the identification of the TT is not as straightforward. Along with the evolution of the PSDs, we investigate evolution of the spectral parameters and the hardness intensity diagram and cite results from previous work on these sources. Below, we explain each case and justification of our TTs. 

 \GX\ in 2003: A detailed analysis of the timing properties of this source is given in \cite{Belloni05}. During the decay, the source enters the hard-intermediate state from the soft state on \wsim\ MJD~52,695. Instead of getting harder monotonically (moving to right in the hardness-intensity diagram), the source goes back and forth in the hardness-intensity diagram, softens, go back to the soft state and then hardens again around MJD~52,718. Starting from this date, the rms amplitude also increases monotonically \citep{Dincer13}; thus, we use MJD~52,717.8 as our TT date.
 
 \FF\ in 2001: There are various features in the PSDs of this outburst throughout the hardness intensity diagram \citep[see][for the evolution of quasi-periodic oscillations]{Rodriguez04}. The earlier part of the decay has PSDs with B type QPOs indicating a soft-intermediate state. Investigating the evolution of PSDs, we recognized that the QPOs disappear at around MJD~51,673, and the power-law flux increases at the same time. Therefore, we chose the TT as MJD~51,674.
 
 \HS\ in 2009: Here, we directly utilized results of \cite{Motta10}, which provide an in-depth analysis of spectral and temporal evolution. Since the shape of the PSD changes, a C-type quasi-periodic oscillation appears and the rms amplitude of variability increases for the observation on MJD~55,016, we set the TT at MJD~55014.7.

 \subsection{Index transition (IT)}
 
\begin{figure}[t]
\plottwo{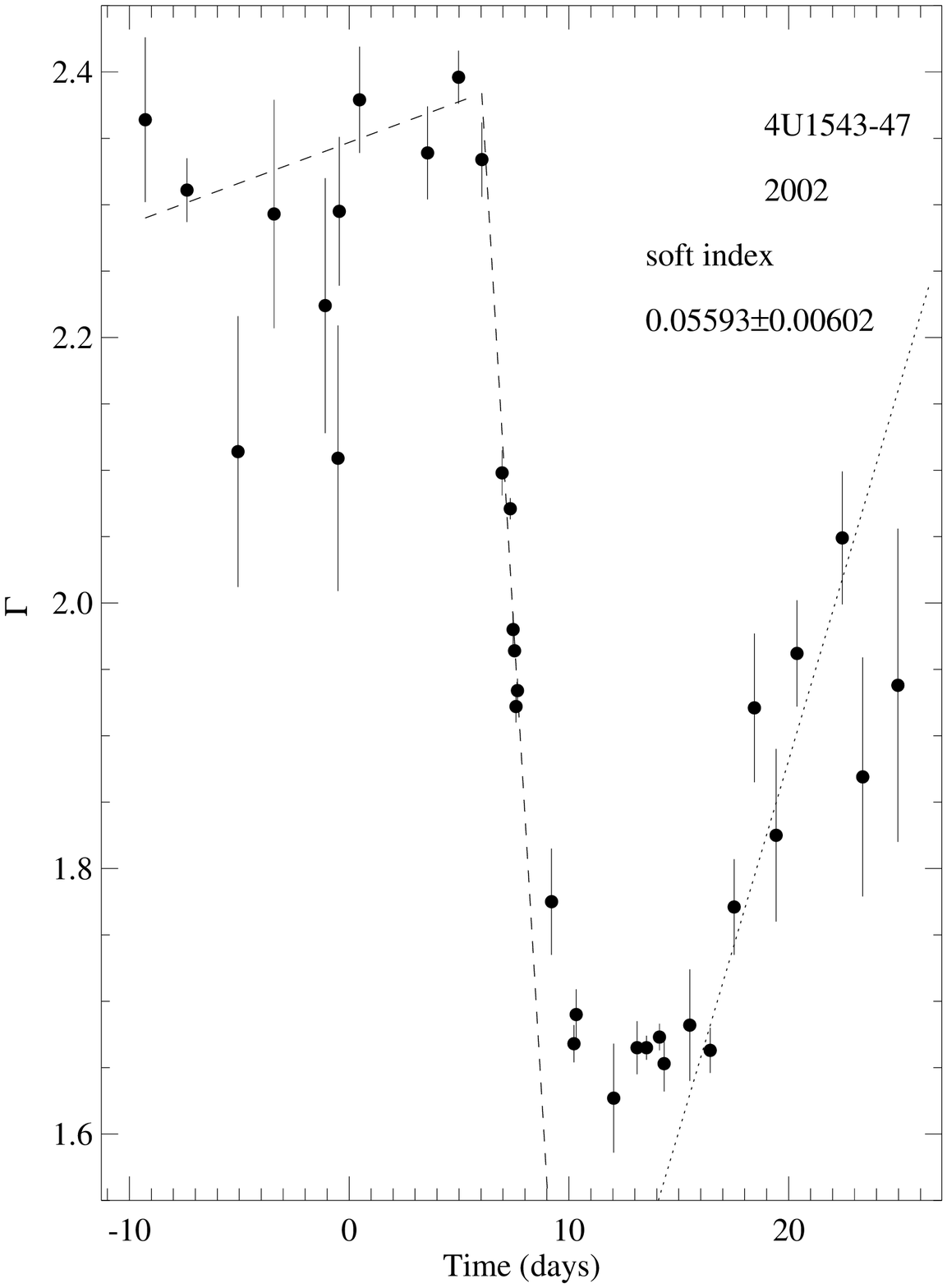}{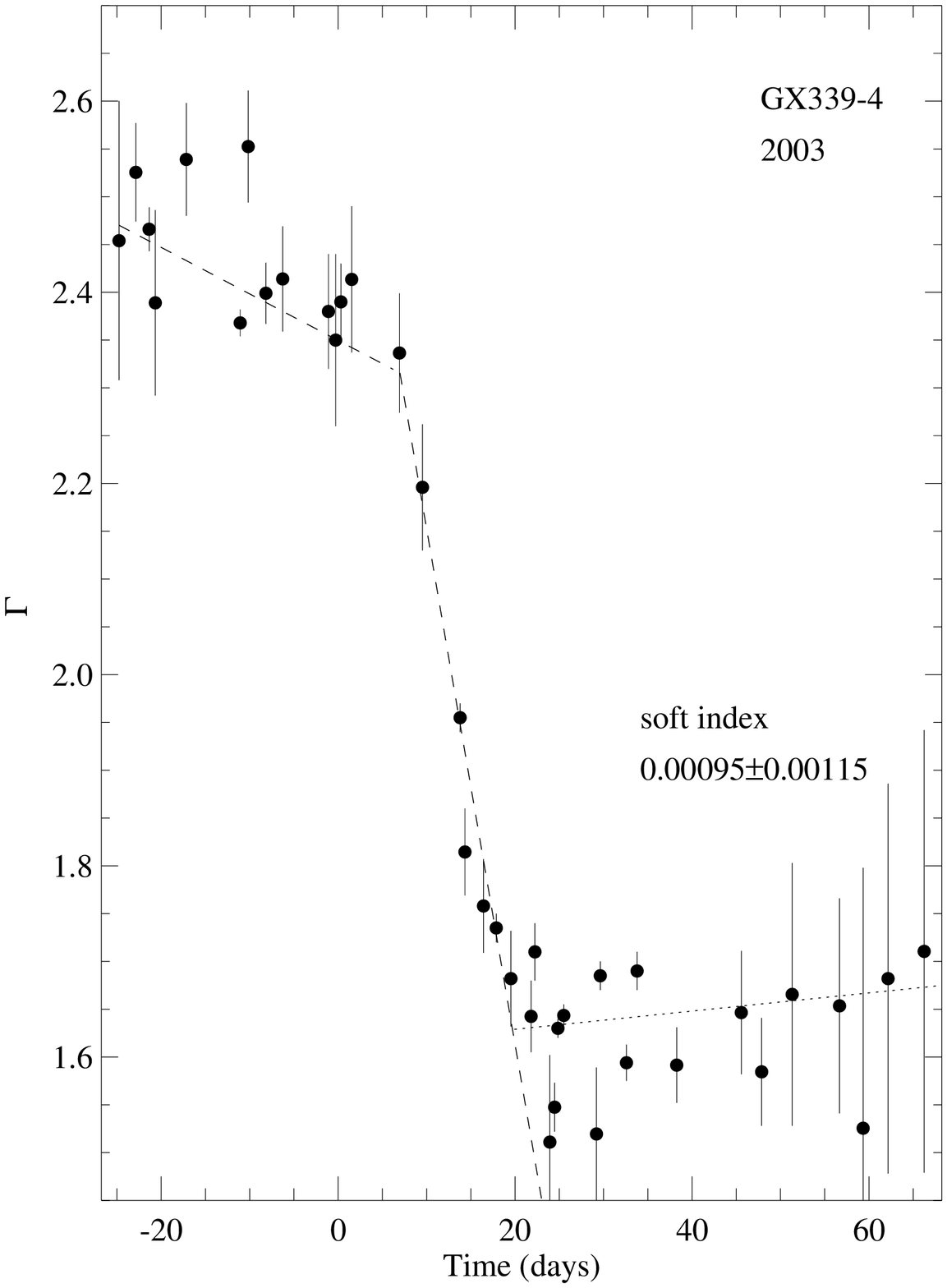}
\caption{\label{fig:indfind}
Left: Evolution of photon indices for \FU\ during the 2002 outburst decay. Fits to the group of observations before hardening, and during hardening are  indicated by the dashed lines. The slope of the fit  for the final 9 observations (dotted lines) clearly indicates softening. Right: Similar fits for the evolution of photon  indices of \GX\ during the 2003 outburst decay. No evidence of softening.
}
\end{figure}

 As seen in Figure~\ref{fig:fuevol} (left) for one source, and in Figure~\ref{fig:ind} for all cases, GBHTs during outburst decay harden quite rapidly from photon index \wsim 2 to \wsim 1.6 over a timescale of \wsim 10 days. To find the IT time, defined as the start of the hardening, we apply two linear fits, one to the group of observations before the apparent drop, and another one to the indices over the time they are dropping (see Figure~\ref{fig:indfind} for two examples). The time corresponding to the intersection of these two lines provides the time of the IT. 
 
 \epsscale{1.1}
\epsscale{1.0}
 
Using the data at the very ends of the outbursts, a third linear fit to the photon index vs. time data of all outburst decays is used to determine if there is softening.  The slope of the line is used as our diagnostic.  For most of the sources, the slopes are consistent with being zero within 1$\sigma$; hence, we conclude that there is no softening for these sources over the range of luminosities being investigated (see Figure~\ref{fig:indfind}, right, for an example). For three sources (\FU, \FF, \SF), the fits result in positive slopes (see Figure~\ref{fig:indfind}, left, for an example). 
 
\subsection{NIR transition}
 
\begin{figure}[t]
\plottwo{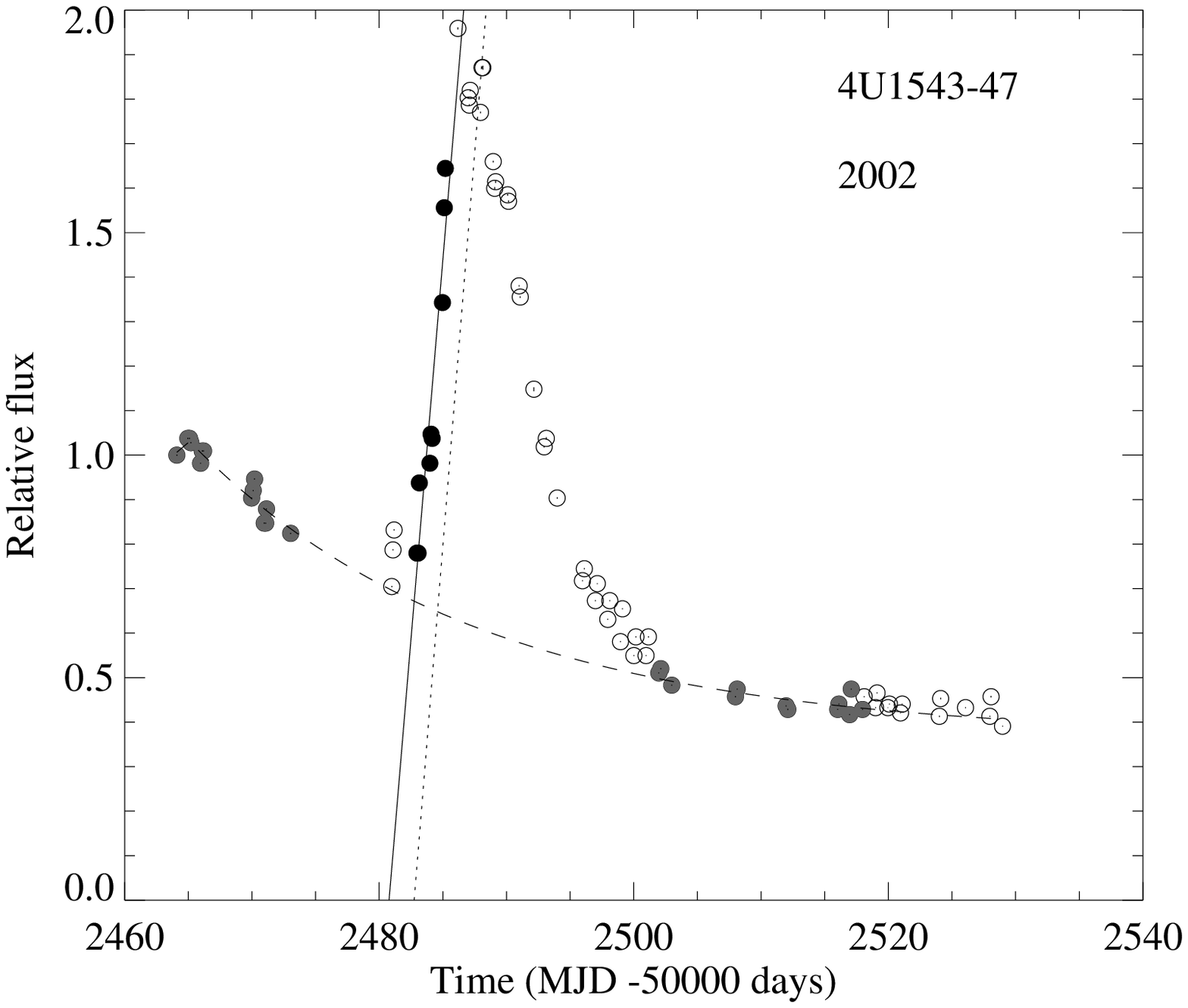}{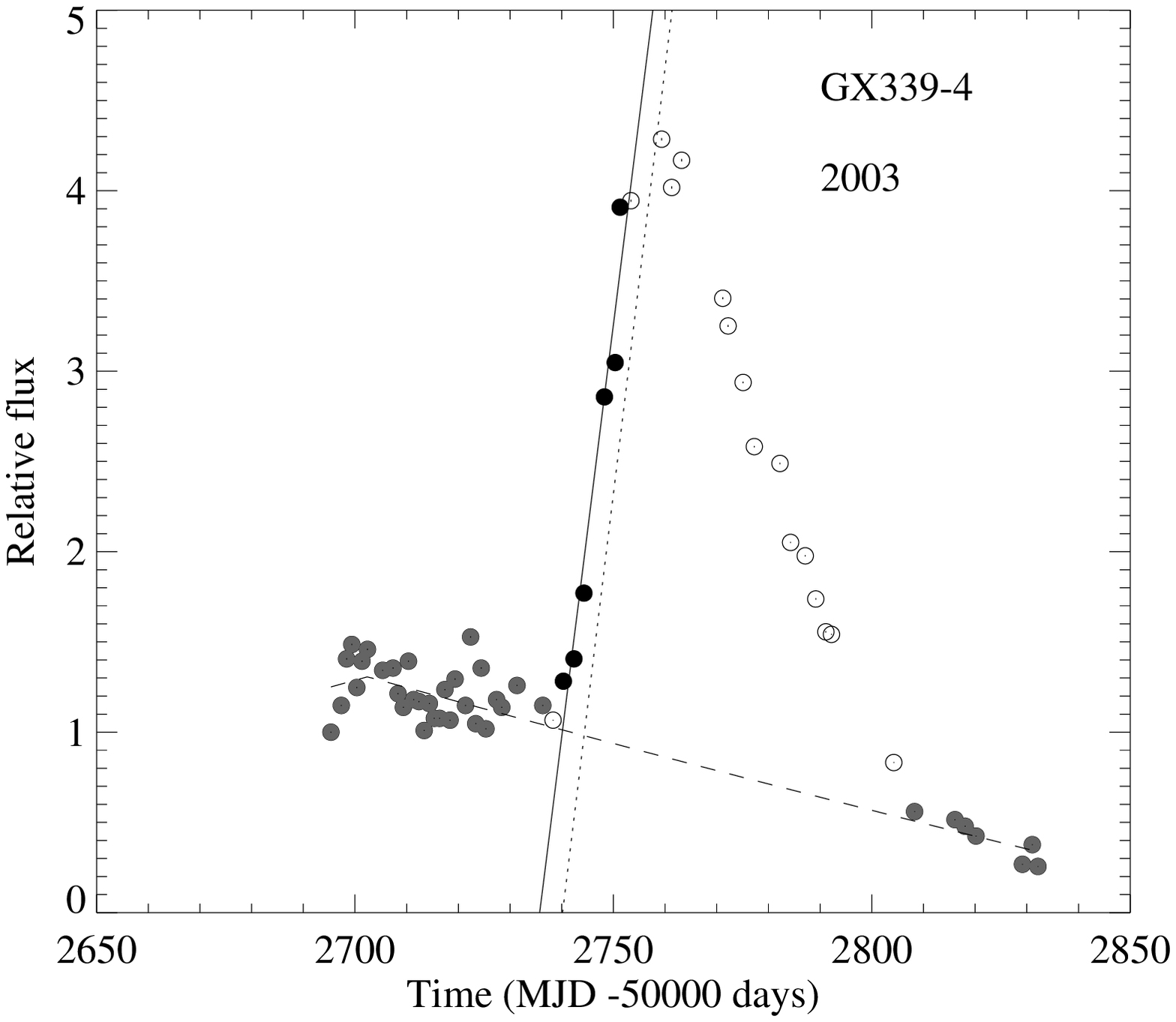}
\caption{\label{fig:irfind}
Left: Evolution of the relative H-band NIR fluxes for \FU\ during the 2002 outburst decay (left), and \GX\ during the 2003 outburst decay (right). For both cases, the gray solid circles represent the baseline, and the black solid circles are the observations that we fit for the NIR rise. The dashed, dotted and the solid lines represent the fit to the baseline, to the NIR rise over the baseline, and to the overall fit; baseline+rise, respectively.
}
\end{figure}

To find the time of the NIR transition, we first convert NIR magnitudes into relative fluxes (compared to the first observation in the dataset). We choose points before and after the flare (gray solid circles in Figure~\ref{fig:irfind}), and apply an exponential fit. These points determine the baseline which has an origin other than the jet, such as the outer parts of the disk or the secondary star (shown with the dashed line in Figure~\ref{fig:irfind}).  We then choose points that represent the rapid rise (excluding the top of the flare, black solid circles in Figure~\ref{fig:irfind}), and fit these points with a linear function over the baseline (dotted line in Figure~\ref{fig:irfind}). The start time is defined as the date that the linear fit intersects zero. This is the same method used in \cite{Kalemci05, Dincer12, Chun13}.

We note that some arbitrariness exists in this method because the points included in the flare and baseline fits are chosen by eye. Also, as discussed in \cite{Buxton12}, the baseline before and after the flare might have an offset, and a single exponential may not fit all points. Despite these potential problems in the method, trying different groups of points for the baseline, omitting the part of the baseline after the flare (e.g., for \GX\ in 2007, the NIR peak does not decay for more than 100 days; therefore, we only used the data before the rise), or using slightly different groups of points for the rise changes the start time at most by a day. 

  \epsscale{1.1}
\epsscale{1.0}

\subsection{Radio Transition}

Here, we provide the details and references for the radio transition times and first radio detections for each source and outburst (see \S~\ref{subsec:multi}, and Table~\ref{tabtr}). 

\GX: For the 2005 outburst, the first radio detection during the outburst decay on MJD~53,481.7 shows an inverted radio spectrum \citep{Corbel12}. Similarly, the radio spectrum is flat on MJD~54,251.7, at the time of first radio detection during the 2007 decay \citep{Corbel12}. Therefore the radio transitions for \GX\ in 2005 and 2007 are also the first time the sources are detected after the TT. For the 2011 decay, the evolution from an optically thin to an optically thick radio spectrum is observed \citep{Corbel13}. The first radio detection is on MJD~55,598.9, and the first optically thick radio detection is on MJD~55,610.1 which correspond to the radio transition.

\FU: The first detection in radio is by MOST on MJD~52,490 at a single frequency. The first multi-frequency radio observation with ATCA is on MJD~52,493 for which the radio spectrum is slightly inverted. 

\SF: On MJD 53,630 and 53,631, the source is not detected with a 4.86 GHz flux density upper limit of 0.3 mJy and a 8.46 GHz flux density of 0.4 mJy. The first detection in radio on MJD 53,634 showed a flat spectrum with $\alpha = 0.27\pm0.76$. 

\ST:  The compact core of this source is detected with the VLBI on MJD~55,311.5 \citep{Yang11}. We set this date as the radio transition since we associate this transition to the presence of a compact jet. Since there are no earlier VLBI observations after the TT, this date should be taken as an upper limit as indicated in Table~\ref{tabtr}. Earlier radio detections exist even at the time of the TT with negative spectral indices indicating optically thin emission \citep{Chun13, Brocksopp13}.

\HS:

2003: The first detection in radio is with the VLA is on MJD~53,940 with  $\alpha = -0.73\pm0.72$. We set the radio transition at the second detection, on MJD~53,948, where the index is $0.82\pm0.87$. 

2008: There are detections of this source with the VLA in radio even before the TT, yet the radio spectra are clearly optically thin (see Fig.~\ref{fig:harden2}, and \citealt{Jonker10}). The single ATCA observation on MJD 54,493, about 5.5 days after the TT, provides an upper limit of 0.15 mJy at 8.46 GHz \citep{Kalemci08_atel}. The first radio detection after the upper limit is on MJD~54,499.7. The radio spectral index becomes consistent with flat within 1 $\sigma$ on MJD~54,502.6, and stays flat afterwards \citep{Jonker10}. Therefore, we place the radio transition at this date.


2009: \cite{MillerJ12} provides information from all the VLA and ATCA radio data of this source for the entire outburst, including a detailed discussion of the jet turn on during the decay. Here, we only show the 4.9 and 8.4 GHz data from VLA along with the evolution of the spectral index in Fig.~\ref{fig:harden2}. Radio emission is detected throughout the soft state with ATCA at a single frequency of 8.4 GHz. The first observation that satisfies our criteria for the radio transition is on MJD~55,026.2, which is also indicated as a transition from the hard intermediate state to low hard state in \cite{MillerJ12}.  

\STT:  The first radio detection of the source during the decay is on MJD 52,728.6, about 2 days after the TT, at a single frequency of 4.86 GHz \citep{Brocksopp05}. The second detection is around 9 days after the TT, again at 4.86 GHz. The first multi-frequency observation is on MJD 52,755.5, 29 days after the TT (see Fig.~\ref{fig:harden2}). The radio spectrum is flat at this date, consistent with emission from a compact jet.

\end{document}